\documentclass[cup7a]{cupbook}
\usepackage{graphicx}
\usepackage{epsfig}
\usepackage{natbib}
\newcommand{\Te}{$T_{\rm{e}}$}

\newcommand{\Hb}{\ifmmode {\rm H}\beta \else H$\beta$\fi}

\title{\it What can emission lines tell us?}
\author[Gra\.zyna Stasi\'nska]{Gra\.zyna Stasi\'nska\\ LUTH, Observatoire de Paris-Meudon\\ 5, Place Jules Janssen, 92195 Meudon Cedex, France \\ 
(grazyna.stasinska@obspm.fr)}

\begin{document}
\pagenumbering{roman}
\maketitle
\tableofcontents
%\cleardoublepage
\pagenumbering{arabic}

\chapter[What can emission lines tell us?]{\it What can emission lines tell us?}

Emission lines are observed almost everywhere in the Universe, from the Earth's atmosphere (see Wyse \& Gilmore 1992 for a summary) to the most distant objects known (quasars and galaxies), on all scales, and at all wavelengths, from the radio domain (e.g. Lobanov 2005) to the gamma rays (e.g. Diehl et al. 2006). They provide very efficient tools to explore the Universe, measure the chemical composition of celestial bodies, and determine the physical conditions prevailing in the regions where they are emitted. 

The subject is extremely vast. In these lectures, we will restrict ourselves in wavelength, being mostly concerned with the optical domain, with some excursions to the infrared and ultraviolet domains and, occasionally, to the X-rays.

We will mainly deal with the mechanisms of line production, and with the interpretation of line intensities in different astrophysical contexts. We will not discuss quasars and Seyfert galaxies, since those are the subject of the lectures by Brad Peterson, nor Lyman $\alpha$ galaxies, which are extensively covered by Daniel Schaerer in this book. However, we will discuss diagnostic diagrams used to distinguish active galaxies from other emission line galaxies, and will mention some topics linked with H Ly$\alpha$.  Most of our examples will be taken from recent literature on planetary nebulae, H\,{\sc ii}  regions and emission line galaxies. Emission line stars are briefly described in the lectures by Stephen Eikenberry in the same volume, and a more detailed presentation is given in the book ``The Astrophysics of Emission Line Stars'' by Kogure \& Leung (2007). 

The vast subject of molecular emission lines has been left aside. The proceedings of the  symposium ``Astrochemistry: Recent Successes and Current Challenges'' (Lis et al. 2006) give a fair introduction to this rapidly expanding field.

In the present text, we will not go into the question of doppler shifts or line profiles, which tell us about radial velocities, and thus about dynamics. This is of course a very important use of emission lines, which would deserve another series of lectures. For example, in such objects as planetary nebulae or supernova remnants, emission line profiles allow one to measure expansion velocities in these objects, and thus investigate their dynamics. The distribution of radial velocities of planetary nebulae in galactic haloes is a way to probe their kinematics and infer the dark matter content  of galaxies (Romanowsky 2006). Redshift surveys  which are used to map the 3D distribution of galaxies in the Universe strongly rely on emission lines (e.g. Lilly et al. 2007), which are the safest tool to measure redshifts. 

We will however mention the great opportunity offered by integral-field spectrosocopy at high spectral resolution, which provides  line intensities and profiles  at every location in a given field of view. With appropriate techniques, this allows one  to recover the 3D geometry of  a nebula.

The purpose of these lectures is not to review all the literature on ionized nebulae, but rather to give clues for understanding the information given by emission  lines, to provide some tools for interpreting one's own data, and to argue for the importance of physical arguments and common sense at each step of the interpretation process.  Therefore, we will review methods rather than objects and papers. This series of lectures complements in some sense the lectures entitled ``Abundance determinations in  H\,{\sc ii}  regions and planetary nebulae'' (Stasi\'nska 2004), where the reader is referred to. In order to save space, the topics that have been extensively treated there will not be repeated, unless we wish to present a different approach or add important new material.

In the following, we will assume that the reader is familiar with Sect. 1 (Basic physics of photoionized nebulae),  Sect. 2 (Basics of abundance determinations in ionized nebulae) and Sect. 3 (Main problems and uncertainties in abundance determinations) of the lectures by Stasi\'nska (2004). We also recommend reading  Ferland's outstanding (2003) review  ``Quantitative spectroscopy of photoionized clouds''. Those wishing a more complete description of the main physical processes occuring in ionized nebulae should consult the text books ``Physical Processes in the Interstellar Medium'' by Spitzer (1978), ``Physics of 
Thermal Gaseous Nebulae'' by Aller (1984), ``Astrophysics of the Diffuse Universe'' by Dopita \& Sutherland (2003), and ``Astrophysics of Gaseous Nebulae and Active Galactic Nuclei'' by Osterbrock \& Ferland (2006). For a recent update on
X-ray astrophysics, a field which  is rapidly developing, one may consult the
AIP Conference Proceedings on 
``X-ray Diagnostics of Astrophysical Plasmas: Theory, Experiment, and Observation'' (2005, Volume 774, edited by Randall K. Smith).

\section{Generalities}

This section is meant to complement the Section ``Basic physics of photoionized nebulae'', which presented the main physical processes occuring in ionized nebulae.  

\subsection{Line production mechanisms}

Emission lines arise in diffuse matter. 
They are produced whenever an excited atom (or ion)  returns to lower-lying levels by emitting discrete photons.  
There are three main mechanisms which produce atoms (ions) in excited levels: recombination, collisional excitation, and photoexcitation.

\subsubsection{Recombination}

Roughly two thirds of the recombinations of an ion occur onto excited states from  which deexcitation proceeds by cascades down to the ground state.
The resulting emission lines are called recombination lines and are labelled with the name of the  \emph {recombined} ion, although their intensities are proportional to the abundance of the \emph {recombining} species. The most famous (and most commonly detected) ones are 
 H\,{\sc i} lines (from the Balmer, Paschen, etc ... series) which arise from recombination of H$^+$ ions; He\,{\sc i}  lines ($\lambda$5876  ...) which arise from recombination of He$^{+}$; and 
He\,{\sc ii}  lines ($\lambda$4686 ...) which arise from recombination of He$^{++}$ ions. Recombination lines from heavier elements are detected as well, (e.g. C\,{\sc ii} $\lambda$4267,  O\,{\sc ii} $\lambda$4651 ...)  but they are  weaker than recombination lines of hydrogen by several orders of magnitude, due to the much  lower abundances of those elements.

The energy $e_{ijl}$ emitted per unit time in a line $l$ due the recombination of the ion $j$ of an element $X^{i}$ can be written as 
\begin{equation}
e_{ijl} =
 n_{e} n(X_{i}^{j}) e^{\rm 0} _{ijl} T_{e}^{-\alpha}, 
\end{equation}
where $e^{\rm 0} _{ijl}$ is a constant  and the exponent $\alpha$ is of the order of 1. Thus, recombination line intensities increase with decreasing temperature, as might be expected. 

\subsubsection{Collisional excitation} 

Collisions with thermal electrons lead to excitation onto levels that are low enough to be attained. Because the lowest lying level of hydrogen is at 10.2~eV, collisional excitation of hydrogen lines is effective only at electron temperatures \Te\ larger than $\sim 2\times 10^4$~K. On the other hand, heavy elements such as nitrogen, oxygen, neon have low lying levels which correspond to fine structure splitting of the ground level. Those can be excited at any temperature that can be encountered in a nebula, giving rise to infrared lines. At ``typical'' nebular temperatures of 8000 -- 12000~K, levels with excitation energies of a few eV can also be excited, giving rise to optical lines. Slightly higher temperatures are needed to excite levels corresponding to ultraviolet lines.

In the simple two-level approximation and when each excitation is followed by a radiative deexcitation, the energy $e_{ijl}$ emitted per unit time in a line $l$ due to collisional excitation of  an ion $j$ of an element $X^{i}$  can be written as
\begin{equation}
e_{ijl}
= n_{e} n(X_{i}^{j}) q_{ijl} h\nu _{l} = 
8.63\,10^{-6}n_{e} n(X_{i}^{j}) \Omega _{ijl}/\omega  _{ijl} T_{e}^{-0.5} 
{\rm e}^{_ (\chi _{ijl}/kT_{e})} h\nu _{l},
\end{equation}
where $\Omega _{ijl}$ is the collision strength, $\omega _{ijl}$ is the 
statistical weight of the upper level, and $\chi _{ijl}$ is the 
excitation energy.

Collisionally excited lines (CELs) are traditionally separated into forbidden, semi-forbidden and permitted lines, according to the type of electronic transition involved. Observable forbidden lines have transition probabilites of the order $10^{-2}$ s$^{-1}$ (or less for infrared lines), semi-forbidden lines of the order of $10^{2}$ s$^{-1}$, and permitted lines of the order of $10^{8}$ s$^{-1}$. This means that the critical density (i.e. the density at which the collisional and radiative deexcitation rates are equal) of forbidden lines is much smaller than that of intercombination lines  and of permitted lines. Table 2 of Rubin (1989) lists critical densities for optical and infrared lines, while Table 1 of Hamman et al. (2001) gives critical densities for ultraviolet lines. 

Resonance lines are special cases of permitted lines: they are the longest wavelenght line arising from ground levels. Examples of resonance lines are  H Ly$\alpha$,  C\,{\sc iv} $\lambda$1550.

\subsubsection{Fluorescent excitation} 

Permitted lines can also be produced by photoexcitation due to stellar light or to nebular recombination lines. The Bowen lines (Bowen 1934) are a particular case of fluorescence, where O\,{\sc iii} is excited by the He\,{\sc ii} Ly$\alpha$ line, and returns to the ground level by cascades  giving rise O\,{\sc iii} $\lambda$3133, 3444, as well as to the line O\,{\sc iii} $\lambda$304 which in turn excites a term of N\,{\sc iii}.

The interpretation of fluorescent lines is complex, and such  lines are not often used for  diagnostics of the nebulae or their ionizing radiation. On the other hand, it is important to know which lines can be affected by fluorescence, in order to avoid improper diagnostics assuming pure recombination (see Escalante \& Morisset 2005). Order of magnitude estimates can be done with the simple approach of Grandi (1976). 

Quantitative analysis of fluorescence lines requires heavy modelling. It can be used to probe the He\,{\sc ii} radiation field in nebulae (Kastner \& Bhatia 1990).  The  fluorescent X-ray iron line has been used to probe accretion flows very close to massive black holes (Fabian et al. 2000).

\subsubsection{Some hints}           
	
Each line can be produced by several processes, but usually there is one which dominates. 
There are, however, cases when secondary processes may not be ignored. For example, the contribution of collisional excitation to 
H Ly$\alpha$ is far from negligible at temperatures of the order of $2 \times 10^4$~K. The contribution of recombination to the intensities of [OII]\,$\lambda\lambda$7320, 7330 is quite important at low temperatures (below, say, 5000\,K) and becomes dominant at the lowest temperatures expected in H\,{\sc ii} regions (see Stasi\'nska 2005).

In the appendix, we give tables of forbidden, semi-forbidden and resonance lines for ions of C, N, O, Ne, S, Cl, Ar that can be observed in H\,{\sc ii} regions and planetary nebulae. These are extracted from the Atomic line list maintained by Peter van Hoof, available at\\http://www.pa.uky.edu/\~{}peter/atomic. This database contains identification of about one million allowed, intercombination and forbidden atomic transitions with wavelengths in the range from 0.5\AA\ to 1000$\mu$m.

W. C. Martin and W. L. Wiese  produced a very useful atomic physics ``compendium of basic ideas, data, notations and formulae''   available at: 
\newline http://www.physics.nist.gov/Pubs/AtSpec/index.html.

\subsubsection{Atomic data}

In the interpretation of emission lines, atomic data play a crucial role. Enormous progress has been done in atomic physics these recent years but not all relevant data are yet available or known with sufficient accuracy. The review by 
Kallman \& Palmeri (2007) is the most recent critical compilation of atomic data for emission line analysis and photoionization modelling of X-ray plasmas. A recent assessment of atomic data for planetary nebulae is given by Bautista (2006).

Many atomic data bases are available on internet. 
\newline - A compilation of data bases for atomic and plasma physics is found at:\newline http://plasma-gate.weizmann.ac.il/DBfAPP.html, 
\newline - Reference data at:\newline http://physics.nist.gov/PhysRefData/\~{}physical, 
\newline - Ultraviolet and X ray radiation at:\newline http://www.arcetri.astro.it/science/chianti/chianti.html, 
\newline - Atomic data for astrophysics (but only up to 2000) at:\newline http://www.pa.uky.edu/\~{}verner/atom.html,
\newline - Atomic data from the Opacity Project at:\newline http://cdsweb.u-strasbg.fr/topbase/topbase.html,  
\newline - Atomic data from the IRON project at:\newline http://cdsweb.u-strasbg.fr/tipbase/home.html.

\subsection{The transfer of radiation}

This section is not to provide a detailed description of radiative transfer techniques, but simply to  mention the problems and the reliability of the methods that are used.

\subsubsection{The transfer of Lyman continuum photons emitted by the ionizing source}

The photons emitted by a source of radiation experience geometrical dilution as they leave the source. They may be absorbed on their way by gas particles, predominantly by hydrogen and helium. The first photons to be absorbed are those which have energies slightly above the ionization threshold, due to the strong dependence of the photoionization cross section with frequency (roughly proportional to $\nu^{-3}$). This gives rise to a ``hardening'' of the radiation field as one approaches the outer edge of an ionized nebula. Photons may also be absorbed (and scattered) by dust grains. 
 
\subsubsection{The transfer of the ionizing photons produced by the nebula}

Recombination produces photons that can in turn ionize the nebular gas.   
These photons are emitted in all directions, so that their transfer is not simple. Authors of photoionized codes have developed several kinds of approximations to deal with this.

The ``on-the-spot approximation'' (or OTS) assumes that all the photons recombining to the ground-state are reabsorbed immediately and at the locus of emission. This is approximately true far from the ionizing source, where the population of neutrals is sufficiently large to ensure immediate reabsorption. However, in the zones of high ionization (or high ``excitation'' as often improperly said) this is not true. Computationally, the OTS assumption allows one to simply discount all the recombinations to the excited levels. This creates a spurious temperature structure in the nebula, with the temperature being overestimated in the high excitation zone, due to the fact that the stellar ionizing radiation field is ``harder'' than the combined stellar $+$ recombination radiation field. The effect is not negligible, about 1000 -- 2000\,K in nebulae of solar chemical composition. Because of this, for some kind of problems, it might be preferable to use a simple 1D photoionization code with reasonable treatment of the diffuse radiation than a 3D code using the OTS approximation. 

The OTS approximation, however, has its utility in dynamical simulations incorporating radiation transfer, because it significantly decreases the computational time. 

Note that the OTS approximation is valid on a global scale. In the integrated volume of a Stromgren sphere, the total number of ionizing photons of the source is exactly balanced by the total number of recombinations to excited levels. This is a useful property for analytic estimations, since it implies that the total H$\alpha$ luminosity of a nebula which absorbs all the ionizing photons (and  is devoid of dust) is simply proportional to $Q_{\rm H}$, the total number of ionizing hydrogen photons\footnote {This property is formally true only for a pure hydrogen nebula, but it so happens that absorption by helium  and subsequent recombination produces line photons that ionize hydrogen and compensate rather well for the photons absorbed by helium}. 

In increasing order of accuracy (and complexity), then comes the ``outward only approximation'', which was first proposed in 1967 by Tarter in his thesis, very early in the era of photoionization codes (see Tarter et al. 1969 for a brief description). Here, the ionizing radiation produced in the nebula is computed at every step, but artificially concentrated in the outward-directed hemisphere, where it is distributed isotropically. This gives a relatively accurate description of the nebular ionizing radiation field, since photons that are emitted inwards tend to travel without being absorbed until they reach the symmetrical point relative to the central source. The great advantage of this approximation is that it allows the computation of a model without having to iterate over the entire volume of the nebula. The code {\sc photo} used by Stasi\'nska, and the code {\sc nebu} used by P\'equignot and by Morisset are based on this approximation. The code {\sc cloudy}, by Ferland, uses the outward-only approximation in a radial-only mode. It appears, from comparisons of benchmark models (e.g. P\'equignot et al. 2001),  that the global results of models made with codes that treat the transfer of diffuse radiation completely (e.g. the code {\sc nebula} by Rubin) are quite similar. Note that the full outward-only approximation allows one to compute the ionizing radiation field in the shadows from optically thick clumps by artificially suppressing the stellar radiation field blocked by the clump.

Codes treating the transfer of ionizing continuum photons exactly, iterating over the entire nebula, interestingly appeared also at the beginning of the era of photoionization codes (Harrington 1968, Rubin 1968). At that time, computers were slow and had little memory, and only spherical or plane parallel geometries could be treated by such codes.

With the present computational capacities, one can do much better and treat the transfer problem accurately for any geometry, by using Monte-Carlo methods. The first such code is {\sc mocassin}, by Ercolano (see Ercolano et al. 2003). One advantage of Monte-Carlo methods is that they allow one to treat the transfer accurately also in extremely dusty nebulae (Ercolano et al. 2005), while the outward-only approximation breaks down in such cases. 

The transfer of resonance line radiation produced in the nebula  is the most difficult to treat accurately, at least in classical approaches of the transfer. This is because it is generally treated in the ``escape probability'' approximation. The effect of line transfer is crucial in optically thick X-ray plasmas such as the central regions of active galactic nuclei (AGN). The code {\sc TITAN} by Dumont treats the transfer of line radiation in an ``exact'' manner, using the ``Accelerated Lambda Iteration'' method (Dumont et al. 2003).

 \subsubsection{The non-ionizing  lines emitted by the nebula}

In general, non-ionizing photons in dust-free nebulae escape as soon as they have been emitted (except perhaps in AGNs were column densities are higher). 
Resonance lines constitute one exception: they may be trapped a long time in the nebula, due to multiple scattering by atoms which, in nebular conditions, are predominantly in their ground level. 

In  dusty  objects, line photons suffer absorption and scattering by dust grains, on their path out of the nebula. Resonance lines, whose path length  can be multiplied by enormous factors due to atomic scattering, are then preferentially affected by dust extinction.  Their observed luminosities then represent only a lower limit to the total energy produced by these lines.

Another case where the diagnostic potential is expected to be reduced due to transfer effects is that of infrared fine structure lines of abundant ions, such as   [O\,{\sc iii}] $\lambda$88, 52 $\mu$m, which can become optically thick in massive  H\,{\sc ii}  regions (Rubin 1978). However, due to a combination of independent reasons, this appears not to be the case even in the extreme situations explored by Abel et al. (2003).

\section{Empirical diagnostics based on emission lines}

 \subsection{Electron temperature and density}
 
 It is well known, and mentioned in all textbooks, that some line ratios  (e.g. the ratios of the lines labeled A1 and  N3 in Table 1.11) are strongly dependent on the temperature, since they have different excitation energies. If the critical densities for collisional deexcitations are larger than the density in the medium under study, these line ratios only depend on the temperature, and are ideal temperature indicators. The most frequently used is the  [O\,{\sc iii}] $\lambda$4363/5007 ratio.
 
 On the other hand, collisionally excited lines that arise from levels of similar excitation energies have their ratios that depend only on the density.  The most common density indicator in the optical is the [S\,{\sc ii}] $\lambda$6716/6731 ratio. Other ones can be easily found by browsing Table 1.11. Rubin (1989) gives a convenient list of optical and infrared line density indicators showing the density range where each of them is useful. 

Similar plasma diagnostics are now available in the X-rays (Porquet \& Dubau 2000, Delahaye et al. 2006, see also Porter \& Ferland 2006).

 \subsection{Ionic and elemental abundances }
 
 There are basically four methods to derive the chemical composition of ionized nebulae.  The first one, generally thought to be the royal way, is through tailored photoionization modelling. The second is by comparison of given objects with a grid of models. These two methods will be discussed in the next section. In this section, which deals with purely empirical methods, we will discuss the other two: direct methods, which  obtain an abundance using information directly from the spectra, and statistical methods, which use relations obtained  from families of objects.

 \subsubsection{Direct methods}
 
 In such methods, one first derives \emph {ionic} abundance ratios directly from observed line ratios of the relevant ions:

 \begin{equation}
\frac{I_{ijl}}{I_{i'j'l'}} =  \frac{ \int{n(X_{i}^{j}) n_{\rm e} \epsilon_{ijl}(T_{\rm e},n_{\rm e})  d V }} {\int{n(X_{i'}^{j'}) n_{\rm e}\epsilon_{i'j'l'}(T_{\rm e},n_{\rm e})  d V }}
, 
\end{equation}
 
where the $I$'s are the intensities and the $\epsilon$'s are given by $e_{ijl}=\epsilon_{ijl} n(X_{i}^{j}) n_{\rm e}$.

Therefore:
 \begin{equation}
\frac{\int{n(X_{i}^{j})} n_{\rm e} dV}{\int{n(X_{i'}^{j'})} n_{\rm e} dV} = \frac{I_{ijl}/I_{i'j'l'}}
{ \epsilon_{ijl}(T_l,n_l) / \epsilon_{i'j'l'}(T_{l'},n_{l'})  }, 
\end{equation}

where $T_l$ and $n_l$ are respectively the electron temperature and density representative of the emission of the line $l$. 

Assuming that the chemical composition is uniform in the nebula, one obtains the element abundance ratios:
 \begin{equation}
\frac{n(X_{i})}{n(X_{i'})} = \frac{I_{ijl}/I_{i'j'l'}}
{ \epsilon_{ijl}(T_l,n_l)/ \epsilon_{i'j'l'}(T_{l'},n_{l'})} \, \, icf, 
\end{equation}

where $icf$ is the ionization correction factor, equal to:
 \begin{equation}
icf = \frac{\int{n(X_{i}^{j})/n(X_{i}) n_{\rm e} dV}}{\int{n(X_{i'}^{j'})/n(X_{i'}) n_{\rm e} dV}}.
\end{equation}

In the case where several ions of a same element are observed, one can use a ``global'' $icf$ adapted to the ions that are observed (e.g. $icf$(O$^+$+ O$^{++}$) for planetary nebulae in which oxygen may be found in higher ionization stages). Note that in H\,{\sc ii} regions (except those ionized by hot Wolf-Rayet stars), $icf$(O$^+$+ O$^{++}$)=1. 

The application of direct methods requires a correct evaluation of the  $T_l$'s and $n_l$'s as well as a good estimate of the ionization correction factor. 

For some ions, the $T_l$'s can be obtained from emission line ratios such as [O\,{\sc iii}] $\lambda$4363/5007 or[N\,{\sc ii}] $\lambda$5755/6584. For the remaining ions, the $T_l$'s are derived using empirical relations with $T(4363/5007)$ or $T(5755/6584 )$ obtained from grids of photoionization models. The most popular empirical relations are those listed by Garnett (1992). A newer set of relations, based on a  grid of models that reproduces the properties of H\,{\sc ii} galaxies, is given by Izotov et al. (2006). It must be noted, however, that observations show larger dispersion about those relations than predicted by photoionization models. It is not clear whether this is due to underestimated observational error bars, or to additional processes not taken into account by photoionization models. At high metallicities \footnote {Throughout the paper the word ``metallicity'' is used with the meaning of  ``oxygen abundance''. This is common practise in nebular studies. Although oxygen is not a metal according to the definition given by chemistry, in nebular astronomy the word metal is often used to refer to any element with atomic weight  $\ge$ 12. The use of the O/H abundance ratio  to represent the ``metallicity'' -- as  first done by Peimbert (1978) -- can be justified by the fact that oxygen represents about half of the total mass of the ``metals'', and that it is the major actor -- after hydrogen and helium -- for the emission spectrum of nebulae. Note that, for stellar astronomers, the word ``metallicity'' is related to the iron abundance, and not to the oxygen abundance, so that the two uses of the word ``metallicity'' are not strictly compatible, since the ratio O/Fe changes during the course of chemical evolution.}, the relevance of any empirical relation between the various $T_l$'s is even more questionable, due to the existence of large temperature gradients in the nebulae, which are strongly dependent on the physical conditions.

The ionization correction factors that are used are either based on ionization potential considerations, or on formulae obtained from grids of photoionization models. For H\,{\sc ii} galaxies, a set of 
$icf$'s is given by Izotov et al. (2006). For planetary nebulae, a popular set of ionization correction factors is that from Kingsburgh \& Barlow (1994), which is based on a handful of unpublished photoionization models. Stasi\'nska (2007, in preparation) gives a set of $icf$s for planetary nebulae based on a full grid of photoionization models. It must be noted, however, that theoretical $icf$'s depend on the model stellar atmospheres that are used in the photoionization models. Despite the tremendous progress in the modelling of stellar atmospheres in the recent years,  it is not yet clear whether predicted spectral energy distributions (SEDs) in the Lyman continuum are correct. 

Finally, note that the line-of-sight ionization structure, in the case of observations that sample only a small fraction of the entire nebula, is different from the integrated ionization structure. This is especially important to keep in mind when dealing with trace ionization stages.

 \paragraph{A  case of failure of $T_e$-based abundances: metal rich giant H\,{\sc ii} regions }
 
 Until recently, it was not possible to measure the electron temperature in metal-rich H\,{\sc ii} regions. The usual temperature diagnostics involve weak auroral lines which easily fall below the detection threshold at low temperatures. With very large telescopes such temperature-sensitive line ratos as  [O\,{\sc iii}] $\lambda$4363/5007, [N\,{\sc ii}] $\lambda$5755/6584, [S\,{\sc iii}] $\lambda$6312/9532 can now be measured even at high metallicities (e.g. Kennicutt et al. 2003, Bresolin et al. 2005, Bresolin 2007). However, due to the large temperature gradients expected to take place in high-metallicity nebulae, which are a consequence of the extremely efficient cooling in the O$^{++}$ zone due to the infrared [O\,{\sc iii}] lines, [O\,{\sc iii}] $\lambda$4363/5007 does not represent the temperatures of the O$^{++}$ zone. As a consequence, the derived abundances can be strongly biased, as shown by Stasi\'nska (2005). The magnitude of the bias depends on the physical properties of the H\,{\sc ii} region and on which observational temperature indicators are available.

 A further problem in the estimation of $T_e$  at high metallicity is the contribution of recombination to the intensities of  collisionally excited lines, which becomes important at low values of $T_{
\rm e}$. For example, the contribution of recombination from O$^{++}$ to the intensity of [O\,{\sc ii}] can be very important.  It can be corrected  for using the formula given in Liu et al. (2000), provided that the temperature characteristic of the emission of the recombination line is known. If the temperature is measured using ratios of CELs only, such is not the case.

\subsubsection{Statistical methods}

 In many cases, the weak [O\,{\sc iii}] $\lambda$4363 or [N\,{\sc ii}] $\lambda$5755 lines are not available because either the temperature is too low, or the spectra are of low signal-to-noise, or else 
the data consist of narrow band images in the strongest lines only.
Then, one may use the so-called ``strong line methods'' to derive abundances. Such methods are only statistical, in the sense that they allow one to derive the metallicity of an H\,{\sc ii} region only on the assumption that this H\,{\sc ii} region shares the same properties as those of the H\,{\sc ii} regions used to calibrate the method. In practise, such methods work rather well for giant H\,{\sc ii} regions, since it appears that giant H\,{\sc ii} regions form a narrow sequence (see e.g. McCall et al. 1985), in which the hardness of the ionizing radiation field and the ionization parameter are closely linked to the metallicity.  Indeed, an increased metallicity enhances the metal line blocking of the emergent stellar flux in the extreme ultraviolet and softens the ionizing spectrum. In addition, the pressure exerted on the nebular gas increases with the strength of the stellar winds, which are related to metallicity, and this in turn decreases the ionization parameter (Dopita et al. 2006).

 Unlike direct methods for abundance determinations, statistical methods have to be calibrated. The reliability of these methods depends not only on the choice of an adequate indicator, but also on the quality of the calibration. This calibration can be done using grids of  ab-initio photoionization models (McGaugh 1991), using a few tailored photoionization models (Pagel et al. 1979), abundances derived from direct methods (Pilyugin \& Thuan 2005), or objects other than H\,{\sc ii} regions thought to have the same chemical composition (Pilyugin 2003)

The oldest and still most popular statistical method is the one based on oxygen lines.  Pagel et al. (1979) introduced the    ([O\,{\sc ii}] $\lambda$3727 +  [O\,{\sc iii}] $\lambda$4959,5007)/H$\beta$ ratio (later referred to as R$_{23}$ or O$_{23}$) to estimate O/H. This method has been calibrated many times,  with results that may differ by about 0.5 dex. 
 Mc Gaugh (1994) and later Pilyugin (2000, 2001)  refined the method to account for  the ionization parameter.  
 
 Many other metallicity indicators have been proposed:   [O\,{\sc iii}] $\lambda$5007/[N\,{\sc ii}] $\lambda$6584 (O$_{3}$N$_{2}$) by Alloin et al. (1979); [N\,{\sc ii}] $\lambda$6584/H$\beta$ (N$_2$), by Storchi-Bergmann et al. (1994) ; ({[S\,{\sc iii}]\,$\lambda$9069}+{[S\,{\sc ii}]\,$\lambda$6716,6731})/H$\alpha$\ (S$_{23}$) by V\'{i}lchez \& Esteban (1996);  [N\,{\sc ii}] $\lambda$6584/[O\,{\sc ii}] $\lambda$3727) (N$_{2}$O$_{2}$) by Dopita et al. (2000); [Ar\,{\sc iii}] $\lambda$7135/[O\,{\sc iii}] $\lambda$5007) (Ar$_{3}$O$_{3}$) and [S\,{\sc iii}] $\lambda$9069/[O\,{\sc iii}] $\lambda$3869) (S$_{3}$O$_{3}$)  by Stasi\'nska (2006);  [Ne\,{\sc iii}] $\lambda$9069/[O\,{\sc ii}] $\lambda$3727) (Ne$_{3}$O$_{2}$)  Nagao et al. (2006).  The metallicity indicators proposed until 2000 have been intercompared by P\'erez-Montero \& D\'{i}az (2005). However, all those methods will have to be recalibrated when the emission line properties  of the most metal rich H\,{\sc ii} regions are well understood, which is not the case at present. 
 
 A few comments are in order. First, any method based on the ratio of an optical  CEL and a recombination line (e.g. O$_{23}$  or S$_{23}$) is bound to be double-valued, as illustrated e.g. by Fig. 7 of Stasi\'nska (2002). This is  because, at low metallicities,  such ratios increase with increasing  metallicity but, at high metallicities, they decrease due to the increased cooling by infrared lines, which lowers the temperature below the excitation threshold of optical CELs. In such circumstances, external arguments must be found to find out whether the object under study is on the ``high abundance'' or ``low abundance'' branch. The most common argument is based on the [N\,{\sc ii}] $\lambda$6584 line. The reason why this argument works is that N/O is observed to increase as O/H increases, at least at high metallicities. Besides, high metallicity H\,{\sc ii} regions tend to have lower ionization parameters, favouring low-excitation lines such as  [N\,{\sc ii}] $\lambda$6584. 
 The biggest problem is at intermediate metallicities, where the maximum of O$_{23}$  and S$_{23}$ occurs, and the metallicity is very ill-determined. By using both O$_{23}$  and S$_{23}$ indexes at the same time, it would perhaps be possible to reduce the uncertainty.  

Methods that use the [N\,{\sc ii}] $\lambda$6584 lines have another potential difficulty. Chemical evolution of galaxies changes the N/O ratio in a complicated and non universal way. Therefore, a calibration is not necessarily relevant for the group of objects under study.

Perhaps the most satisfactory methods, on the theoretical side, are the ones using the Ar$_{3}$O$_{3}$ or S$_{3}$O$_{3}$  indicators, since these indicators are monotonic and work for well understood reasons, directly linked to metallicity. 

On the contrary, the Ne$_{3}$O$_{2}$ index, which is seen to decrease as metallicity decreases, behaves in such a way \emph {only} because more metal rich giant H\,{\sc ii} regions happen to be excited by a softer radiation field and have a lower ionization parameter. This is a very indirect metallicity indicator!

It is important to be aware that, in principle, strong line methods can be safely used only when applied to the same category of objects that were used for the calibration. The meaning of the results in the case of integrated spectra of galaxies, for example, is far from obvious in an absolute sense. Such spectra contain the light from H\,{\sc ii} regions differing in chemical composition and extinction as well as the light from the diffuse ionized interstellar medium. In addition, inclination effects may be important. A few studies have adressed these issues from an observational point of view (Zaritsky et al. 1994, Kobulnicky et al. 1999, Moustakas \& Kennicutt 2006), but clearly the subject is not closed.

A further step in strong line abundance determinations has been made by using ratios of line equivalent widths instead of intensities (Kobulnicky et al. 2003). The advantage of using  equivalent widths is that they are almost insensitive to interstellar reddening, which allows one to apply the method even when  reddening corrections are not available, especially at redshifts larger than 1.6. The reason why equivalent widths work well for integrated spectra of galaxies is that there is empirically a very close correlation between line intensities and equivalent widths, meaning that, statistically, stellar and nebular properties as well as the reddening are closely interrelated.

\subsection{Estimation of the effective temperature of the ionizing stars }
 
 \paragraph{$T_{\star}$ from the Zanstra method.}

This method, proposed by Zanstra (1931), makes use of the fact that the number of stellar quanta in the Lyman continuum, normalized to the stellar luminosity at a given wavelength, is an increasing function of the effective temperature. This is illustrated in Fig. 1.1 (based on modern stellar model atmospheres). In practise, it is the luminosity of the H$\beta$ line which is the counter of Lyman continuum photons. This assumes that all the Lyman continuum photons are absorbed by hydrogen. This assumption breaks down in the case of density bounded nebulae or of nebulae containing dust mixed with the ionized gas.  In real nebulae, some Lyman continuum photons are also absorbed by He$^0$ and He$^+$. However, recombination of these ions produce photons that are able to ionize hydrogen, so that the basic assumption of the Zanstra method is generally remarkably well fulfilled. Of course, the value of the effective temperature, $T_{\star}$, obtained by the Zanstra method will depend on the model atmosphere used in the derivation.

  \begin{figure}
    \centering
%    \vspace{4cm}
\centerline{
\includegraphics[scale=0.17]{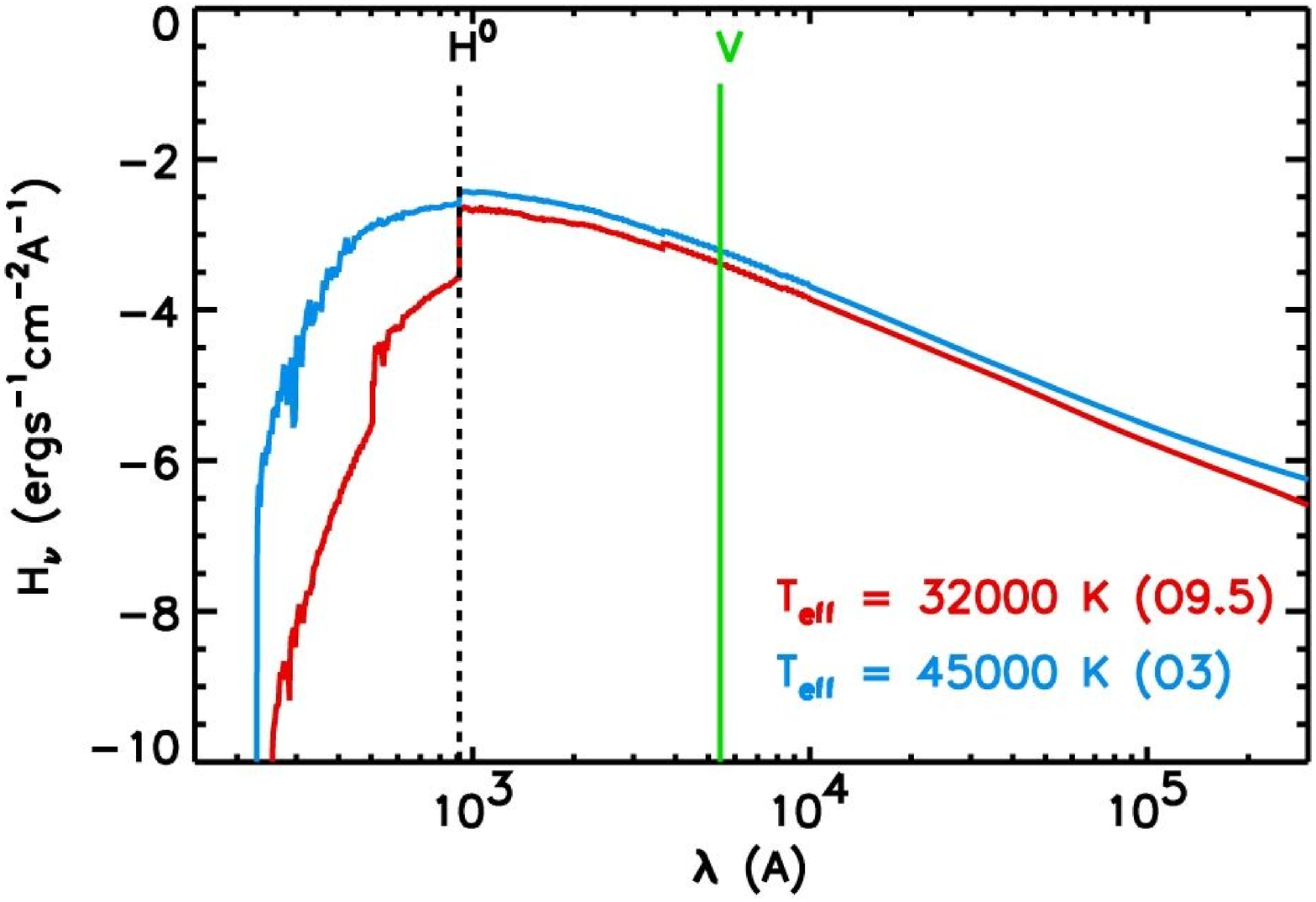}
\includegraphics[scale=0.435]{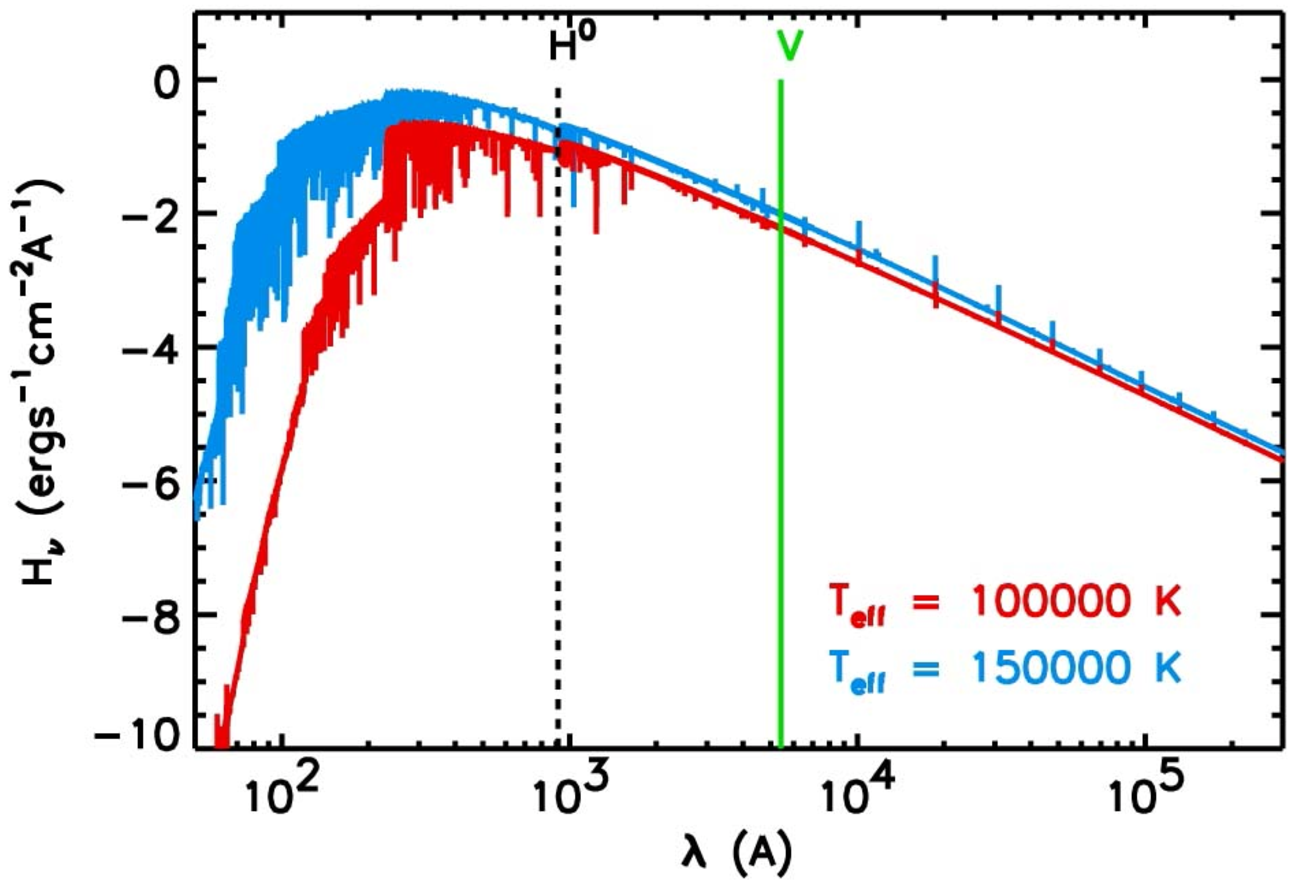}
} 
    \caption{Spectral energy distributions (SEDs) for  effective temperatures corresponding to massive stars (left) and central stars of planetary nebulae (right). The dotted line indicates the position of the ionization potential of hydrogen, the full line indicates the wavelength of the V filter.}
    \label{sample-figure}
  \end{figure}

For very hot stars, such as the central stars of planetary nebulae, one can also define a He$^+$ Zanstra temperature, using  the  He\,{\sc ii} $\lambda$4686 flux as a measure of the number of photons with energies above 54.4 eV.

  \begin{figure}
    \centering
%    \vspace{4cm}
\includegraphics[scale=0.6]{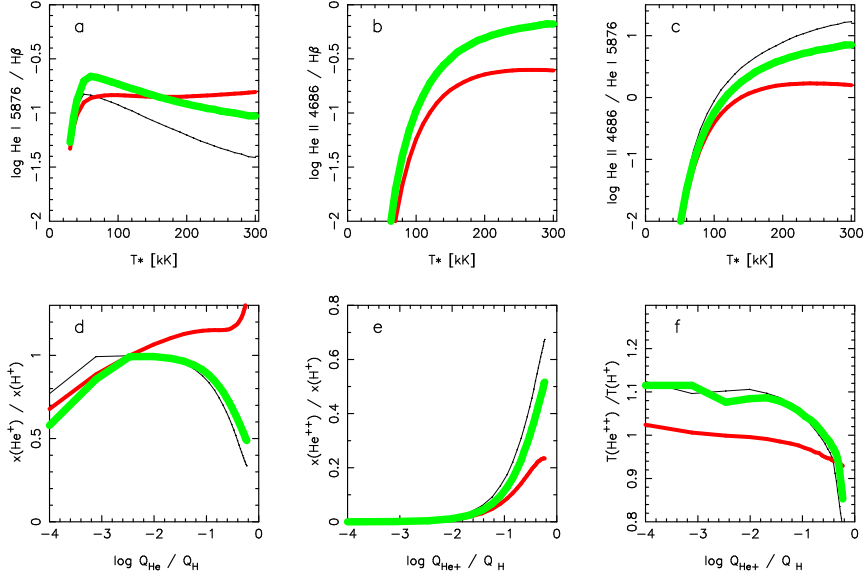}
\caption{Sequences of photoionization models for spherical, uniform nebulae ionized by stars of different effective temperatures. Thin black curve:  ionisation parameter $U$=10$^{-2}$,  He/H=0.1.  Thicker red curve:  $U$=10$^{-3}$, He/H=0.1. Thick green curve: $U$=10$^{-2}$, He/H=0.15. For simplicity, the models were run with a blackbody central star.}
    \label{sample-figure}
  \end{figure}

A few properties, illustrated by the models shown in Fig. 1.2, are worth of noting.  \begin{itemize}
\item He\,{\sc i} $\lambda$5876/H$\beta$ measures $T_{\star}$ only in a small range  of temperatures ($T_{\star}$ $<$ 40 kK for our models with a blackbody  as seen in Fig. 1.2.a, this limit being of course different for realistic stellar atmospheres). 
\item Due to the competition between H$^0$ and He$^+$ to absorb photons with energies above 54.4 eV, He\,{\sc ii} $\lambda$4686/H$\beta$ saturates at $T_{\star}$ $>$ 150 kK and depends on $U$ at $T_{\star}$ $>$ 100 kK (Fig. 1.2.b).
\item He\,{\sc ii} $\lambda$4686/H$\beta$ is independent of He/H (Fig. 1.2.b), while
He\,{\sc ii} $\lambda$4686/He\,{\sc i} $\lambda$5876 depends on He/H (Fig. 1.2.c). Indeed, He\,{\sc ii} $\lambda$4686 is a counter of photons, as long as the He$^{++}$ Stromgren sphere is smaller than the H$^+$ Stromgren sphere.
\item The temperatures in the H$^0$ and He$^+$ zones are not equal (Fig. 1.2.	f). 
\end{itemize}

 \paragraph{$T_{\star}$ from the observed ionization structure.}

Since the ionization structure of a nebula depends on $T_{\star}$,  one can think of using  the line ratios of two successive ions  to infer $T_{\star}$ (e.g. Kunze et al. 1996). However, in such an approach, the effect of the  ionization parameter must be considered as well.

To alleviate this problem, V\'ilchez \& Pagel (1988)
introduced of a ``radiation softness parameter'', which they defined as: 
 \begin{equation}
  \eta = \frac {{\rm O}^{+}/{\rm O}^{++}} {{\rm S}^{+}/{\rm S}^{++}},
   \end{equation}
and expected to be independent of $U$ in first approximation. This, however, was not sufficient for an accurate determination of $T_{\star}$ (or even for ranking the effective temperatures of different objects).

Morisset (2004) constructed a grid of photoionization models with various values of $U$ (at  various metallicities) to allow a proper estimate of $T_{\star}$. The grid was constructed using the {\sc WM-Basic} model atmospheres of Pauldrach et al. (2001).

\paragraph{$T_{\star}$ from energy-balance methods.}

This method was first proposed by Stoy (1933). It makes use of the fact that the heating rate of a nebula is a function of the effective temperature. Since, in thermal equilibrium, the heating rate is compensated by the cooling rate, a measure of the cooling rate allows one to estimate $T_{\star}$. Most of the cooling is done through collisionally excited lines. Therefore, an estimation of $T_{\star}$ can be obtained from the formula: 
\begin{equation}
\sum L_{\rm CEL} / L({\rm H}\beta) = f(T_{\star}).	
\end{equation}
Pottasch \& Preite-Martinez (1983) proposed a calibration of this method for planetary nebulae.

A similar argument led 
Stasi\'nska (1980) to propose a diagram to estimate the average effective temperature of the ionizing stars of giant H\,{\sc ii} regions. This diagram plots the value of  [O\,{\sc iii}] $\lambda$4363/5007 as a function of the metallicity O/H, for photoionization models corresponding to various values of $T_{\star}$.

 \subsection{Determining the star formation rate}

The star formation rate is an important quantity to measure in galaxies. It can be obtained in many ways: ultraviolet continuum, far infrared continuum, radio continuum, recombination lines, forbidden lines. The last two are sensitive to the most recent star formation (less that a few Myr, which is the lifetime of the most massive stars). Note that, while the luminosity in a line measures the absolute value of the recent star formation rate, the equivalent width of the line measures the ratio of the present to past star formation rate.
Any technique must be calibrated using simulations of stellar populations in which the basic parameters are   the stellar initial mass function,  the star formation history, and the metallicity. These simulations are based on libraries of stellar evolutionary tracks and of stellar atmospheres. 
 Kennicutt (1998) and Schaerer (2000) give exhaustive reviews on the question. Here, we simply mention a few issues regarding the estimation of star formation rates using emission lines.

\paragraph{Star formation rate using H$\beta$ or H$\alpha$. }

As any determination of a physical parameter using H$\beta$, the basic assumption is that all the Lyman continuum photons are absorbed by the gas. Also, the effect of intervening dust extinction  must be properly accounted for (see Kewley et al. 2002). As shown by Schaerer (2000), the derived star formation rate strongly depends on the adopted stellar initial mass function and upper stellar mass limit. 

\paragraph{Star formation rate using [O\,{\sc ii}] $\lambda$3727.}
It may seem strange to use such a line to measure the star formation rate, instead of simply using  H$\alpha$. In addition to the provisos on H$\alpha$, the [O\,{\sc ii}] $\lambda$3727 obviously must depend on the metallicity and ionization parameter. The main reason for attempting to use the [O\,{\sc ii}] $\lambda$3727 line in spite of this is that it can be observed in the optical range at larger redshifts than H$\alpha$ or H$\beta$.  However, it is an observational fact that the [O\,{\sc ii}] $\lambda$3727/H$\beta$ ratio strongly varies among emission line galaxies, even discarding objects containing an active nucleus (see Fig. 1.3). Therefore, the use of  [O\,{\sc ii}] $\lambda$3727 as a star formation rate indicator is extremely risky. 

   \begin{figure}
    \centering
\includegraphics[scale=0.6]{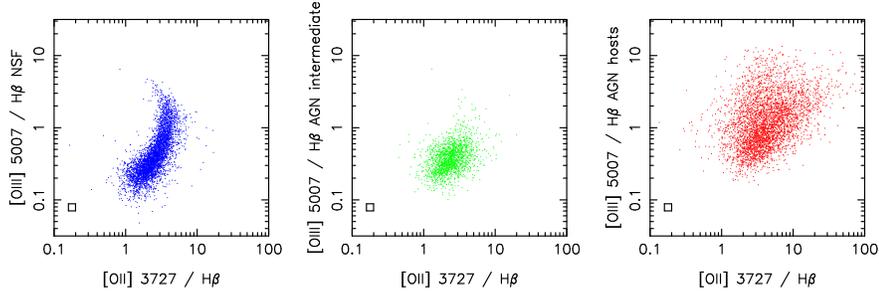}
    \caption{Diagrams using data on galaxies from the Sloan Digital Sky Survey (taken from Stasi\'nska  2006) to show how strongly the [O\,{\sc ii}]\,$\lambda$3727/H$\beta$ ratio varies among normal star forming galaxies (left), galaxies containing an active nucleus (right), and hybrid galaxies (middle). }
    \label{sample-figure}
  \end{figure}

 \subsection{How to distinguish normal galaxies from AGN hosts?}
 
After the discovery of spiral galaxies with a very bright nucleus emitting strong and broad (several thousands of km s$^{-1}$) emission lines (Seyfert 1941), it became clear that these galactic nuclei{\footnote {The term Seyfert galaxy was used for the first  time by  de Vaucouleurs (1960).} were the locus of violent, non-stellar activity (Burbidge et al. 1963, Osterbrock \& Parker 1965), perhaps of the same nature as found in quasars. Heckman (1980) performed a spectroscopic survey of the nuclei of a complete sample of 90 galaxies, and found that the presence of
low ionization nuclear emission-line regions (LINERs) were quite common, and seemed to be the scaled down version of Seyfert nuclei. Baldwin et al. (1981) were the first to propose spectroscopic diagnostics based on emission line ratios to distinguish normal star forming galaxies from  active galactic nuclei (AGN). The most famous is the [O\,{\sc iii}] $\lambda$5007/H$\beta$ vs  [N\,{\sc ii}] $\lambda$6584/H$\alpha$ diagram, often referred to as the BPT diagram (for 
Baldwin, Phillips, \& Terlevich). The physics  underlying such a diagram is that photons from  AGNs are harder than those from the massive stars that power H\,{\sc ii} regions. Therefore they induce more heating, implying that optical collisionally excited lines will be brighter with respect to recombination lines than in the case of ionization by massive stars only. Veilleux \& Osterbrock (1987) proposed additional diagrams: 
[O\,{\sc iii}] $\lambda$5007/H$\beta$ vs  [S\,{\sc ii}] $\lambda$6725/H$\alpha$, and  [O\,{\sc iii}] $\lambda$5007/H$\beta$ vs  [O\,{\sc i}] $\lambda$6300/H$\alpha$. As previously found by McCall et al. (1985),  giant H\,{\sc ii} regions form a very narrow sequence in these diagrams. 
 
 It was a great surprise, after the first thousands of galaxy spectra from the Sloan Digital Sky Survey (SDSS, York et al. 2000) were released, to find that a proper subtraction of the stellar continuum in galaxies (Kauffmann et al. 2003) allowed one to see a second sequence in the BPT diagram, in the direction opposite to that of the star-forming sequence. Thus, emission line galaxies in the BPT diagrams are distributed in two wings, which look like the wings of a flying seagull (see Fig. 1.4).

   \begin{figure}
\includegraphics[scale=0.65]{Stasinska-fig4a.eps}
\includegraphics[scale=0.1]{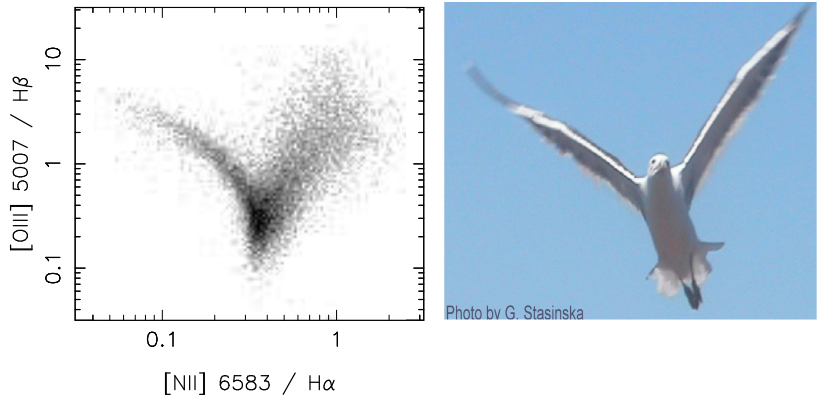}
    \caption{Galaxies from the Sloan Digital Sky Survey in the BPT diagram are distributed in two wings: the wings of a flying seagull. }
    \label{sample-figure}
  \end{figure}

Just a few years before, Kewley et al. (2001) had constructed a  grid of photoionization models  in order to determine a theoretical upper limit to the ionization by massive stars in the BPT diagram. This upper limit, later referred to as the ``Kewley line'' proved well to the right of the star forming wing from the SDSS.  Kauffman et al. (2003) shifted this line to the left to define  an empirical limit between normal star forming galaxies and AGN hosts (the ``Kauffmann'' line). Stasi\'nska et al. (2006) found this limit to be still too ``generous'', and proposed a more restrictive one, based on a grid of photoionization models aimed at reproducing the upper envelope of the left wing of the seagull. Its equation is:

\begin{equation}
y =
(-30.787+1.1358x+0.27297x^2){ \rm tanh}(5.7409x) -31.093,
\end{equation}

  where $y$ =
log [O\,{\sc iii}] $\lambda$5007/H$\beta$ and $x$ =  log [N\,{\sc ii}] $\lambda$6584/H$\alpha$. 

Note that, among the three diagnostic diagrams proposed by Veilleux and Osterbrock (1989), the [O\,{\sc iii}] $\lambda$5007/H$\beta$ vs  [N\,{\sc ii}] $\lambda$6584/H$\alpha$ is the most efficient, due to the increase of N/O in galaxies of high metallicity. Then, heating by an AGN boosts the [N\,{\sc ii}] line and creates a clear separation of the two wings. One then understands why the [O\,{\sc iii}] $\lambda$5007/H$\beta$ vs  [O\,{\sc ii}] $\lambda$3727/H$\beta$ diagram, used as a surrogate of the BPT diagram at redshifts of $\sim$ 0.2 (Lamareille et al. 2004) is much less efficient in separating AGN hosts from normal star forming galaxies.

Stasi\'nska et al. (2006) also noted that a classification based on [N\,{\sc ii}] $\lambda$6584/H$\alpha$ only is also feasible, and has the advantage of being applicable to a much larger number of galaxies, since [O\,{\sc iii}] $\lambda$5007 and H$\beta$  are not needed. 

 Finally, these authors proposed another classification diagram, plotting $D_n(4000)$\footnote {$D_n(4000)$ is the discontinuity observed at 4000\AA\ in the spectra of galaxies; it increases with stellar metallicity and age.} as a function of the equivalent width of [O\,{\sc ii}] $\lambda$3727 (the $DEW$ diagram). The rationale for this diagram is that if in a galaxy with a large $D_n(4000)$ emission lines can be seen, they must be due to another cause than photoionization by massive stars. The hope is to be able to use this diagram to distinguish AGNs at redshifts higher than 0.2. However, the fact that higher redshifts correspond to younger ages of the stellar populations is a issue that requires further investigation. 
 
To end up this section, let us remark that the term ``LINER'' is nowadays often employed to designate those galaxies that are found to the lower right of the BPT diagram. This is an unfortunate deviation from the original meaning, since it is likely that, in those galaxies, the emission lines do not come only from the nucleus, but from a much larger zone of the galaxy. The original ``LINERS'' discovered by Heckman would not appear in the BPT diagram due to the low luminosity of the active nucleus, unless there is significant line emission from the rest of the galaxy.  

\section{Photoionization modelling}

We now turn to the ``royal way'' of analysing emission line spectra: photoionization modelling. We will show that photoionization modelling is an art which not only requires photoionization codes, but also a certain dose of common sense. 

 \subsection{Publicly available photoionization codes} 
 
Most of the codes listed below have been intercompared at the Lexingtion conference ``Spectroscopic Challenges of Photoionized Plasmas'' in 2000. The results of the code comparisons can be found   in P\'equignot et al. 2001.
 
\subsubsection{1D photoionization codes}

 \begin{itemize}
 \item 
 {\sc cloudy} by Gary Ferland and associates, 
 computes models for ionized nebulae and photo-dissociation regions (PDR). It is regularly updated, well documented, and widely used. It is available at \newline
 http://www.nublado.org
 \item {\sc mappings} by Michael Dopita  + Kewley, Evans, Groves, Sutherland, Binette, Allen, Leitherer 
 computes models for photoionized nebulae and for planar shocks. It can be found at  \newline
http://www.ifa.hawaii.edu/\~{}kewley/Mappings
 \item {\sc Xstar} by Tim Kallman  computes models for photoionized regions with special attention to the treatment of X-rays. It is found at  \newline
 http://heasarc.nasa.gov/lheasoft/xstar/xstar.html  
 \end{itemize}
 
 \subsubsection{3D photoionization codes}
 
  \begin{itemize}
 \item {\sc cloudy}-3D by Christophe Morisset  is a pseudo 3D code, based on  {\sc cloudy}. It allows  quick modelling of 3D nebulae and visualization 
 (including computation and visualization of line profiles). It can be found at \newline
 http://132.248.1.102/Cloudy\_3D
\item  {\sc mocassin} by Barbara Ercolano is a full 3D Monte-Carlo photoionization code that also treat dust transfer in an accurate manner. It is available from \newline be@star.ucl.ac.uk (see also http://hea-www.harvard.edu/\~{}bercolano)  
 \end{itemize}

 \subsubsection{Other codes}
Many other, independent photoionization codes are mentioned in the literature (some of them  benchmarked in P\'equignot et al. 2001), but are not of public access. This is especially the case of hydrodynamical codes that include the physics of photoionization and the computation of line emission. In such codes, of course, the physics of radiation is treated in a simplified manner, since a simultaneous  treatment of the macro- and microphysics requires tremendous computing power.

 \subsection{Why do photoionization models?}
There can be many different reasons for which to build photoionization models. For example, one might want to: 
\begin{itemize}
 \item Check the sensitivity of input parameters on observable properties.
 \item Compute a grid of models for easy interpretation of a certain class of objects.
 \item Calculate ionization correction factors.
 \item Derive the chemical composition of a given nebula.
 \item Estimate characteristics of the ionizing source.
 \item Probe stellar  atmosphere model predictions in the far ultraviolet.
 \end{itemize}

  \subsection{How to proceed?}
  
Each of the problems above requires a specific approach. It is always worth spending some time to find the best way to achieve one's goal. For example, if one wants to derive the chemical composition of a nebula by means of emission line fitting, it is not sufficient to find one solution. One must explore the entire range of possible solutions, given the observational constraints. This is not always easy.

If the aim is to interpret one object, or a given class of objects, the first step is to collect  all the observational constraints needed for this purpose. This includes monochromatic images as well as line intensities in various wavelengths ranges and in different apertures. Also it is important to characterize the ionizing sources as best as possible from the observations (visual magnitude, spectral type in the case of a single star, age of the ionizing stellar population in the case of a large collection of coeval stars).

Then, one must define a strategy: 
How to explore a parameter space?
How to deal with error bars?
How to test the validity of a model?

Finally, one must evaluate the result of the investigation: 
Was the goal achieved? If  not, what does this imply?  In fact, this last aspect is often overlooked, while it is potentially very instructive and is an incentive for progress in the field. 

In the following we present some commented examples.

  \subsection{Abundance derivation by tailored model fitting}
  
The general procedure would be something like:

1) Define the input parameters: 
The characteristics of the ionizing radiation field (luminosity, spectral energy distribution); 
The density distribution of the nebular gas; 
The chemical composition of the nebular gas;  
The distance.

2) Use an appropriate photoionization code and compute a model.

3) Compare the outputs  with the observations (corrected for extinction by intervening dust):
The total observed H$\alpha$ flux;
The H$\alpha$ surface brightness distribution and the angular size of H$\alpha$ emitting zone;
The line intensities etc...

4) Go back to 1) and iterate until the observations are satisfactorily reproduced. Here ``satisfactorily'' means that \emph {all} the observational data are reproduced within acceptable limits, those limits taking into account both the observational errors and the approximations of the model. These ``limits'' should be set by a critical analysis of the situation, \emph {before} running the models. It may be that one finds no solution. This is by no means a defeat. It is actually an important result too, which tells something about the physics of the object. But, in order to be useful, such a result must be convincing, in other words the reason why no solution can be found must be clearly explained.
\hspace{1 pt}

For a good quality model fitting, it is important to:

1) Use as many observational constraints as possible 
(not only line intensity ratios).

2) Keep in mind that the importance of the constraint has nothing to do with the strengths of the lines. 
For example He\,{\sc ii} $\lambda$4686/H$\beta$, which is of the order of a few percent at most in H\,{\sc ii} regions,  indicates the presence of photons above 54.4 eV, which are not expected in main sequence, massive stars (unless they are part of a  X-ray binary system); [O\,{\sc iii}] $\lambda$4363/5007, [N\,{\sc iii}] $\lambda$5755/6584 indicate whether the energy budget is well reproduced. It is too often ignored that photoionization models do not only compute the ionization structure of the nebulae, they also compute their temperature, which is essential in predicting the strengths of the emission lines.

3) Recall that some constraints are not independent. For example, 
if [O\,{\sc iii}] $\lambda$5007/H$\beta$ is fitted by the model, then [O\,{\sc iii}] $\lambda$4959/H$\beta$ should be fitted as well because the [O\,{\sc iii}] $\lambda$5007/4959 ratio is fixed by atomic physics, since both lines originate from the same level\footnote {The  lack of variation of the  [O\,{\sc iii}] $\lambda$5007/4959 ratio among 
quasars at various redshifts shows that the fine structure constant, $\alpha$, does not depend on cosmic time (Bahcall et al. 2003). A far-reaching application of emission-line astrophysics!}. In no case can the fact that both lines are fitted at the same time be taken as a success of the model. 
On the other hand, if they are not, this may indicate an observational problem, e.g. that the strong [O\,{\sc iii}] $\lambda$5007 line is saturated. As a matter of fact, many observers precisely use this ratio as a check of the accuracy of their observations.

4) Choose a good estimator for the ``goodness of fit'', 
e.g. avoid using a $\chi^2$ minimization technique without being convinced that this is the most appropriate test in the case under study.
All the observables should be fitted (within limits defined a priori).

5) Try to visualize the comparison model-result as much as possible. Examples among many possibilities can be found in Stasi\'nska \& Schaerer (1999) or in Stasi\'nska et al. (2004).

\paragraph {Outcomes from model-fitting:}

A priori, the most satisfactory situation is when \emph {all} the observations are fitted within the error bars.
This may imply (but not necessarily) that the model abundances are the real abundances. If the constraints are insufficient,
the model abundances may actually be very different from the true ones.

Quite often,  some of the observations cannot be fitted.
This means  either that the observations are not as good as thought, or that the model does not represent the object well.    
Some assumptions in the modelling may be  incorrect. For example the nebula has a different geometry than assumed, or the stellar ionizing radiation field
is not well described, or an important heating mechanism is missing. 
In such a situation,  the chemical composition is generally not known with the desired accuracy, a fact too often overlooked.

\paragraph {An example of photoionization modelling without a satisfactory solution: }The most metal-poor galaxy I ZW 18
  (Stasi\'nska \& Schaerer 1999)}

 The observational constraints were provided by  
 narrow band imaging in nebular lines and stellar continuum, along with 
 optical spectra giving nebular line intensities. The 
 ionizing radiation field was given by  a  stellar population synthesis  model aimed at reproducing the observed features from hot Wolf-Rayet stars. 
The conclusion from the modelling exercise was that, even taking into account strong deviations from the adopted spectral energy distribution of the ionizing radiation and the effect of postulated additional X-rays, the photoionization models yield too low a  [O\,{\sc iii}] $\lambda$4363/5007 ratio by about 30\%\footnote {Note that the observed geometrical constraint provided by the images was important in reaching this conclusion, since the authors also showed that assuming a simpler geometry with no central hole -- but in complete disagreement with the observations -- one could obtain a higher electron temperature, due to reduced H Ly$\alpha$ cooling.}. This significant discrepancy   cannot be solved by expected inaccuracies in the atomic data. The missing energy is of the same order of magnitude as the one provided by the stellar photons.

Interestingly, this paper was rejected by a first referee on the account that ``the temperature of the nebula, as indicated by the
ratio [O\,{\sc iii}] $\lambda$4363/5007, is in all trial models
out of the observed range by 30$\%$.
The authors seem to have tried everything possible [...]  but still ... no secure results.'' A second referee accepted the paper without changes emphasizing  that ``The
results are of highest interest for future studies of ionized regions,
mainly by the identification of a serious problem in the interpretation of
the [O\,{\sc iii}] $\lambda$4363/5007 ratio, which cannot be consistently reproduced by the
models ''!

\paragraph {Another example of photoionization modelling without a satisfactory solution:} The giant H\,{\sc ii} region NGC\,588 in the galaxy M33
  (Jamet et al. 2005).

In this case, the observational constraints were provided by a detailed 
characterization of the ionizing stars (Jamet et al. 2004),
narrow band imaging,
long slit optical spectra, and 
far infrared line fluxes from the ISO satellite.  
In the analysis, aperture effects were properly taken into account. The temperature derived from the [O\,{\sc iii}] $\lambda$4363/5007 ratio is larger than the one derived from the 
[O\,{\sc iii}] $\lambda$88$\mu$/5007  ratio  by 3000K. It was found that no photoionization model could explain the observed temperature diagnostics.  Unless the measured [O\,{\sc iii}] $\lambda$88$\mu$ flux is in error by a factor 2, 
this implies that the oxygen abundance in this object is uncertain by about a factor of two.

Note that both this study and the former one used a 1D photoionization code, while the study of the nebular surface brightness distribution indicates that the nebula is not spherical. The nebular geometry was taken into account in the discussion, and the conclusions reached in both cases are robust.

\paragraph {An example of photoionization modelling with insufficient constraints: } The chemical composition of the Galactic bulge planetary nebula     M 2-5. 

This is an object for which  no direct temperature diagnostic was available. Using photoionization modelling,  Ratag (1992, thesis) and Ratag et al. (1997) claimed an oxygen abundance of one fourth solar.  However, as shown in Stasi\'nska (2002), models fulfilling \emph {all} the observational constraints  can be constructed with oxygen abundances as different as O/H= 1.2$\times 10^{-3}$ and 2.4$\times 10^{-4}$! This is linked to the ``double value problem'' discussed in  Sect 1.2.2.2. Because of only crude knowledge of the physics of nebulae and insufficient exploration of the parameter space, Ratag had reached a possibly erroneous conclusion.

    \subsection{Abundance derivation using grids of models}

One is often interested in determining the metallicities of a large sample of objects. The extreme case being emission line galaxies from the Sloan Digital Sky Survey, which are counted by tens of thousands. A  tailored model fitting is out of reach in such cases. Apart from statistical methods, discussed in Sect. 1.2.2.2, one may consider building a vast, finely meshed grid of photoionization models and use them to derive the metallicities of observed emission line galaxies either by interpolation or by Bayesian methods. Such  an approach has been used by Charlot \& Longhetti (2000), or Brinchmann et al. (2004). This method is powerful and appealing, since it can be completely automatized. However, it is not the ultimate answer to the problem of abundance determination and it must be used wih some circumspection. As  shown by Yin et al. (2007),  when applied to a sample of galaxies in which the [O\,{\sc iii}] $\lambda$4363 line is observed, allowing direct abundance determinations, this methods returns significantly higher metallicities than the direct method. Yin et al.  argue that the reason lies in the abundances chosen in the  model grid. Since the procedure uses simultaneously all the strong lines to derive O/H, any offset between the real N/O and the N/O adopted in the model grid induces an offset in the derived O/H. 

  \subsection{Testing  model atmospheres of massive stars using H\,{\sc ii} regions}
  
In an on-going study of the  Galactic H\,{\sc ii} region M\,43 and  its ionizing star (Sim\'on-D{\'{\i}}az et al., in preparation), one of the aims is to use the nebula to probe the spectral energy distribution modelled for its ionizing star. The nebula  has a rather simple structure for an H\,{\sc ii} region, being apparently round. The  characteristics of the ionizing star (effective temperature and gravity) were obtained by fitting the stellar H and He optical lines using the stellar atmosphere code {\sc fastwind} (Puls et al. 2005). The stellar luminosity was then obtained, knowing the distance. The predicted stellar spectral energy distribution was used as an input to the photoionization code {\sc cloudy} (great care was taken in the treatment of the absorption edges, which are not the same in {\sc cloudy} and {\sc fastwind}). The nebula was assumed spherically symmetric, the  density distribution was 
chosen to reproduce the observed H$\alpha$ surface brightness distribution. The starting 
nebular abundances were obtained  using the classical $T_{\rm e}$-based method. The 
observational constraints are the
dereddened line intensities 
 at various positions in the nebula.

 It was found that the distribution of emission line ratios across the nebula 
cannot be reproduced by the model. While this could imply that 
the tested model atmosphere does not predict the energy distribution correctly, another hypothesis must be investigated first: that the geometry assumed for the nebula is wrong. In spite of its roundish appearance, it is possible that M\,43 is  not a sphere, but a blister  seen face on. The constructed photoionization model would then correspond to its  ``spherical impostor'', using the expression 
by  Morisset et al. (2005). As shown by Morisset \& Stasi\'nska (2006), observation of the emission line profiles would allow one to distinguish between the two geometries. In the meantime, photoionization computations using the pseudo 3D code {\sc cloudy}\_3D of Morisset (2006) is being carried out, to provide models for the blister geometry.

  \subsection{Photoionization study of an aspherical nebula using a 3D code: the planetary nebula NGC\,7009}
  
The use of a 3D photoionization modelling is generally difficult, due to the number of parameters involved to describe the geometrical structure of a nebula and to the long computational time required for a model. A 1D code is often sufficient to identify a physical problem that needs solution, as seen in Sect. 1.3.4. On the other hand, a 3D code is essential when tackling a problem where the effect of geometry is important. Such is the case, for example, of the low-ionization knots observed in planetary nebulae, for which it has been said that they show nitrogen enhancement by factors 2--5 (e.g. Balick et al. 1994). Subsequent studies argued that this conclusion might be wrong, due to the use of an improper ionization correction scheme (it was assumed that N/O = N$^+$/O$^+$). Gon{\c c}alves et al. (2006) examined this problem in the case of  NGC\,7009, a planetary nebula with two conspicuous symmetrical  knots. Their aim was to 
explore the possibility that the enhanced [N\,{\sc ii}]  emission observed in the  knots could be due to ionization effects, by building a 3D photoionization model of homogeneous chemical composition reproducing the observed geometry and spectroscopic pecularities. They simulated observations in different regions of the nebula, and found that the N$^+$/O$^+$ ratio varies strongly with position and can explain (at least partly) the apparent nitrogen enhancement in the knots.

  \subsection{The interpretation of data from integral field spectroscopy}
  
 Integral field spectroscopy is becoming a major tool to study nebulae and galaxies. A wealth of spatially resolved data will routinely become available. The question now is which tools to use to interpret such data. Obviously, starting by building 3D photoionization models is not a good way to proceed. As argued above, 3D modelling is difficult and time consuming, and a proper evaluation of the benefits of this kind of approach is recommended before  such an endeavour. On the other side, one may be tempted to use results from grids of published 1D photoionizaion models or observational diagnostic diagrams relevant for the integrated nebular light. This is risky. For example, one might attribute the increase of  [N\,{\sc ii}]/H$\alpha$ or  [S\,{\sc ii}]/H$\alpha$ in certain zones to non-stellar ionization while it might merely be a line-of-sight effect in a region close to an ionization front.

\section{Pending questions}

There are a few important aspects which have not been mentioned above, for the sake of a more linear presentation. Most of them have been extensively treated  in Stasi\'nska (2004), and will be mentioned here only briefly, with some updates if necessary. 

\subsection{Correction for reddening, underlying stellar absorption and aperture effects}

Before being analyzed in terms of abundances or star formation rates or being compared to the results of photoionization models, the intensities of observed lines must be corrected for various effects. The presence of dust between the zone of emission and the observer attenuates the collected radiation and modifies its colour. It appears that the dust extinction curve, which was once considered universal, is actually a one-parameter function characterized by the value of the total-to-selective extinction $R_V$ = $A_V/E(B-V)$ (Fitzpatrick 1999, 2004). The canonical value of $R_V$, generally used for extinction corrections, is  3.1 or 3.2. However, the  measurement of $R_V$ using 258 Galactic O stars yields a distribution around this value with a dispersion of $\pm$ 0.5.  Values of $R_V$ as small as 1.6 or as large as 5 are found (Patriarchi et al. 2003).

The extinction curve of  a
single star by interstellar dust is different from
the obscuration curve of an extended nebula (Calzetti 2001), because in the latter case the observed radiation includes some scattered light.   From a sample of  starburst galaxies observed from the far ultraviolet to the near infrared,  an obscuration curve for galaxies was derived (Calzetti 1997, Calzetti et al. 2000). 

  \begin{figure}
\includegraphics[scale=0.6]{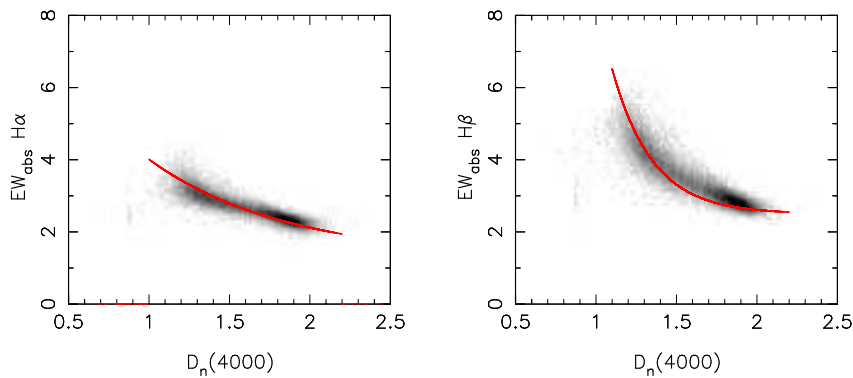}
    \caption{The stellar absorption equivalent withs at H$\alpha$ (left) and H$\beta$ (right) as a function of the discontinuity at 4000\AA.  They have been determined in a sample of galaxies extracted from the Digital Sky Survey by stellar population analysis using the method of Cid-Fernandes et al. (2005). The red curves correspond to the fits given by Eqs. 1.10 and 1.11.}
    \label{sample-figure}
  \end{figure}

The amount of reddening is usually determined by fitting the observed Balmer decrement to the theoretical one at the appropriate temperature. In emission line galaxies, in giant H\,{\sc ii} regions (and also in unresolved planetary nebulae), the observed Balmer lines are affected by the underlying assumption from the stellar component. This absorption can be determined empirically together with the extinction by fitting several Balmer lines to the theoretical Balmer decrement (Izotov et al. 1994). It can also be modelled, using spectral synthesis techniques to reproduce the stellar absorption features other than Balmer lines, and reading out from the model spectrum that best fits the data what is the stellar absorption at the the Balmer lines wavelengths (Cid Fernandes et al. 2005). This procedure applied to a sample of 20000 galaxies from the Sloan Digital Sky Survey (see Stasi\'nska et al. 2006 for a description of the sample) shows a spectacular correlation between absorption equivalent widths at the Balmer line wavelengths and the discontinuity at 4000\AA, $D_n$(4000), as shown for the first time here (Fig. 1.5). Since $D_n$(4000) is relatively easy to measure, this provides an empirical way to estimate the underlying stellar absorption at H$\alpha$ and H$\beta$. The data shown in Fig. 5 can be modelled by the expressions:

\begin {equation}
EW_{\rm {abs}}({\rm H}\alpha) = 9 \exp (-1.2x) + 1.3
\end {equation}

and 

\begin {equation}
EW_{\rm {abs}}({\rm H}\beta) = 400 \exp (-4.x +0.5) + 2.5,
\end {equation}
where $x=D_n(4000)$.

Another issue is when the objects under study are more extended than the observing beam. This fact  must be taken into account in the analysis. For example, spectra obtained with  several instruments must be carefully combined. This is especially important -- and difficult -- when merging ultraviolet or infrared data with optical ones, since in that case flux calibration problems may arise. In case of observations where a large number of lines is available, it is theoretically possible to combine all the data by comparing each observation to the results of a tailored photoionization model seen through an appropriate aperture. Most of the time, however, aperture corrections are performed in an approximate way, neglecting the effect of the ionization structure of the nebulae on the line fluxes recorded through small apertures.

A further problem, when dealing with spectra of entire galaxies is the physical meaning of the derived abundance, especially in the presence of abundance gradients. This issue, which is also relevant to the simple  ``strong line indicators'' approach, has not yet been fully addressed.

\subsection{Escape of ionizing radiation }

Most nebular studies assume that the nebulae are ionization bounded. This affects such issues as the estimate of stellar temperatures via the Zanstra method, the estimation of star formation rates in galaxies, or the outcome of tailored photoionization modelling.

There is, however, growing evidence that many nebulae are density bounded 
in at least some directions. This was known for a long time for planetary nebulae. It is now recognized to be the case also for many giant H\,{\sc ii} regions (Beckman et al. 2002, Stasi\'nska \& Izotov 2003, Castellanos et al. 2003).

\subsection{The importance of stellar energy distribution}

The outcome of photoionization models depends critically on the adopted ionizing SED, especially as regards the nebular ionization structure. There are two main issues in this respect. One is the description of the ionizing spectral energy distribution of an individual star. The other is how one deals with the case of ionization by a group of stars.

\paragraph{Model atmospheres for hot stars.}

The atmospheres of stars are generally extended and experience important deviations from LTE. Hundreds of thousands of metallic lines produce line blanketing and line blocking effects, and are at the origin of radiative driven winds. Modern stellar atmosphere codes are able to handle these aspects, although each one works  with some degree of approximation at least in one aspect. 

The most broadely used in photoionization modelling of H\,{\sc ii} regions are {\sc tlusty} (Hubeny \& Lanz 1995),   {\sc WM-basic} (Pauldrach et al. (2001), and {\sc cmfgen} (Hillier \& Miller 1998). What matters for the modelling of ionized nebulae is the SED. There are sizeable differences among the  predictions from these codes in the ionizing continuum. It is not clear, at present, which model predictions are best suited for photoionization modelling. A few studies (e.g. Morisset et al. 2004a and 2004b) have used observed H\,{\sc ii} region to test the predictions of the model atmospheres. However, this is a difficult task, and the answer is not clear yet. Sim{\'o}n-D{\'{\i}}az et al. (2007) are attacking this problem anew, with both a theoretical and an observational approach. 

The central stars of planetary nebulae can reach much higher effective temperatures than massive stars. Temperatures of over $2 \times 10^5$/,K have been found. When they are close to becoming white dwarfs, these stars have tiny atmospheres for which the plane-parallel approximation is reasonable. A grid of plane-parallel line blanketed  NLTE atmospheres has been computed by Rauch (2003) using the code {\sc pro2} (Werner \& Dreizler 1993). At earlier stages of the planetary nebulae,  the effect of winds are expected to be significant. They are important for planetary nebulae central stars  of [WR] type, which share the same atmospheric properties as massive Wolf-Rayet stars. The same kind of approach as for H\,{\sc ii} regions has been used for planetary nebulae to test the ionizing SEDs predicted by model atmospheres (e.g. Stasi\'nska et al. 2004 who used the Potsdam  code for expanding atmospheres in non-LTE described in Hamann \&
Koesterke 1998).

Some grids of model atmospheres computed with the above codes are available, for example at the {\sc cloudy} website.

\paragraph{Ionization by a group of stars.}

In the case of giant H\,{\sc ii} regions, where the ionization is provided by many stars, one generally relies on SEDs predicted by stellar population synthesis codes.  Starburst 99 (Leitherer 1999, Smith et al. 2002) or  {\sc pegase} (Fioc \& Rocca-Volmerange 1997, 1999) are public-access codes able to do this. 
The integrated SED of an entire stellar population is obviously quite different from the SED of the component stars, but the quality of the SED for photoionization studies is related to the reliability of the predictions in the Lyman continuum of the stellar atmosphere codes used to compute the stellar libraries. 

There is, however, another important issue, pointed out  by Cervi\~no et al. (2000). It is the fact that in real H\,{\sc ii} regions, statistical sampling strongly affects the ionizing SED, and SEDs predicted by traditional stellar population synthesis models may be far from real. Practical considerations as regards the outcome from photoionization models can be found in Cervi\~no et al. (2003) and a detailed discussion on uncertainties in stellar synthesis models and ``survival strategies'' is given by Cervi\~no \& Luridiana (2005). 

One of the yet unsolved problems is the question of the  often observed He\,{\sc ii} emission in H\,{\sc ii} galaxies. Traditional models (Stasi\'nska \& Izotov 2003) have difficulties in accounting for it properly. While there are several possible explanations (X-ray binaries, X-rays from hot star atmospheres, etc...), sampling effects may also be an issue that needs to be considered.

Finally, in the case of a group of ionizing stars, one might wonder if the effect on the H\,{\sc ii} region strongly depends on whether the stars are concentrated or whether they are distributed in the nebula. This question is being adressed by Ercolano et al. (2007)

\subsection{Dust }

The question of the effect of dust in ionized nebulae has been amply reviewed in Stasi\'nska (2004). Among publicly available photoionization codes, {\sc cloudy}, {\sc mappings} and {\sc mocassin} include a treatment of dust grains coupled with the gas particles. 

When it comes to entire galaxies, 
a proper  treatment of dust should take into account the large scale geometrical distribution of dust  inside the galaxies. Charlot \& Longhetti (2001)  did a first step by including  a different prescription  for the
absorption of photons from H\,{\sc ii} regions and from older stars, as 
proposed  by Charlot \& Fall (2000). Panuzzo et al. (2003) implemented results from photoionization simulations with  {\sc cloudy} into models of dusty galaxies computed with the code {\sc Grasil2} (Silva et al. 1998), which include a sound treatment of all the aspects
of dust reprocessing, providing a more realistic description of the integrated spectrum of a galaxy. 

\subsection{Temperature fluctuations and the ORL/CEL discrepancy}

 Since the seminal work by Peimbert (1967), there exists the suspicion that elemental abundances derived from optical collisional emission lines (CELs) might be plagued by the presence of temperature fluctuations in nebulae.  If such temperature fluctuations exist, then the highest temperature zones will favour the lines of higher excitation, and as a result the temperature derived from  [O\,{\sc iii}] $\lambda$4363/5007 will overestimate the temperature of the O$^{++}$ zone, and result in an  understimate of  elemental abundances with respect to hydrogen. 
 
 Fourty years after,  nebular specialists (see e.g. the recent reviews by Esteban 2002, Liu 2002, Ferland 2003, Peimbert \& Peimbert 2003) are still debating about this issue, which is of fundamental importance for the reliability of abundance determinations of nebulae. We are talking here of an effect of typically 0.2 dex in metal abundances relative to hydrogen. One can summarize the debate in four questions: 

1) Is there evidence for the existence of temperature fluctuations at a level that significantly biases the abundances from CELs?

2) If yes, is there a reliable procedure to correct for this bias?

3) What would be the possible causes of such temperature fluctuations?

4) Is the problem the same in H\,{\sc ii} regions and in planetary nebulae?

 At present, there are  no generally accepted answers to these questions. In the following, we simply collect some elements for the discussion, by supplementing what can be found in the above reviews and in Stasi\'nska (2004).

1) Direct evidence for temperature fluctuations high enough to be of sizeable consequence for abundance determinations is scarce. It is provided by Hubble Space Telescope temperature mapping  of the Orion nebula (O'Dell et al. 2003) or spectroscopy (Rubin et al. 2003). Indirect evidence comes from the comparison of temperatures measured by various means (such as [O\,{\sc iii}] $\lambda$4363/5007 and the Balmer or Paschen jump). There are many examples, both in H\,{\sc ii} regions and planetary nebulae, which point at a typical r.m.s. temperature fluctuation, $t^2$, of 0.04 (see references in the review quoted above). However, recent determination of Balmer jump temperatures in H\,{\sc ii} galaxies (Guseva et al. 2007) give  $t^2$ close to zero, on average. Note that these latter measurements concern metal poor H\,{\sc ii} regions. Another piece of evidence comes from the discrepancy of abundances derived from ORLs and CELs for the same ion. In this case, the implied  $t^2$ vary considerably, from less that 0.02 to almost 0.1, with the values for H\,{\sc ii} regions being small, and those for planetary nebulae spanning a wide range.

2) Peimbert (1967) and Peimbert \& Costero (1969) devised a scheme, based on Taylor series expansion of the analytic expressions giving the emissivities, to correct for the abundance bias, with the help of several simplifying assumptions. It can, however, be shown, on the basis of a two-zone toy model (Stasi\'nska 2002), that a proper description of temperature fluctuations and any correction scheme for abundance determinations require at least three parameters, not just two as in Peimbert \& Costero (1969). Besides, due to the restricted number of observational constraints, the application of the Peimbert \& Costero scheme implies some assumptions which are incorrect, like the equality of $t^2$(O$^{++}$)  and $t^2$(H$^{+}$).

3) Some possible causes for the presence of temperature fluctuations have been listed by Torres-Peimbert \& Peimbert (2003) and Peimbert \& Peimbert (2006). These include shocks, magnetic reconnection, dust heating, etc... However, the energy requirements to obtain $t^2$ values of about  0.04 are very large, typically a factor 2 with respect to the energy gains from gas photoionization. Note that it is not so difficult to obtain a $t^2$ of 0.02-0.04 for the entire nebular volume, since the main cooling lines change across the nebulae. What is difficult is to obtain such a large $t^2$ in the O$^{++}$ or O$^{+}$ zones only, which is what would be implied by the ORL/CEL discrepancy. Using a ``hot spot representation'' of temperature fluctuations, Binette et al. (2001)  propose a sound quantification of energy requirements for the observed $t^2$ and find that ``the combined mechanical luminosity of stellar winds, champagne 
flows and photoevaporation 
flows from proplyds is insufficient to account for the temperature 
fluctuations of typical H\,{\sc ii} regions.'' Note that, at high metallicities, the natural temperature gradients in H\,{\sc ii} regions easily lead to a formal value of  $t^2$(O$^{++}$) of 0.04 or even more, because of the effect of strong [O\,{\sc iii}] $\lambda$88, 52$\mu$m cooling. The difficulty is to obtain such a large $t^2$(O$^{++}$) at metallicities smaller than 12+log (O/H) = 8.7.

4) There is now a large data base of high quality observations of metal recombination lines  (see references in Liu 2006 for planetary nebulae and in Garc\'ia-Rojas \& Esteban 2006 for H\,{\sc ii} regions). Analyzing the behaviour of the abundance discrepancy factors in both types of objects  Garc\'ia-Rojas \& Esteban (2006) conclude that the source for the ORL/CEL abundance discrepancy must be different in H\,{\sc ii} regions and in planetary nebulae. Since the paper by Liu et al. (2000), it has been argued that, in planetary nebulae, the ORL/CEL abundance discrepancy is likely due to the presence of hydrogen-poor inclusions in the body of the nebula (see e.g. Liu 2002, 2003, 2006, Tsamis et al. 2004).  These metal-rich clumps would be too cool to excite the CELs. Quite a small mass of matter in the  clumps is sufficient to reproduce the observed ORL/CEL discrepancies. The nature of these clumps could be  photoevaporating planetesimals. Tsamis et al. (2003) and Tsamis \& P\'equignot (2005) suggest that the ORL/CEL discrepancy in H\,{\sc ii} regions might also be due to hydrogen poor inclusions (obviously of different astrophysical origin). This point of view is not universally accepted (see Garc\'ia-Rojas \& Esteban 2006 for a different opinion). Yet, it seems that the galaxy chemical enrichment scenario  proposed by Tenorio-Tagle (1996), in which the oxygen-rich products from supernova explosions end up in metal rich droplets falling back onto the galaxies after an excursion in the intergalactic medium, would be able to explain quantitatively the ORL/CEL abundance discrepancies observed in H\,{\sc ii} regions (Stasi\'nska et al. 2007).

The debate on temperature fluctuations and ORL/CEL discrepancies is far from closed. Yet its outcome will have important implications on the use of emission lines in the derivation of elemental abundances in galaxies. Peimbert et al. (2006) are already proposing to use recombination line abundances to calibrate the strong line methods for abundance derivations. They argue that recombination line abundances are not affected by any temperature bias, which is true in the case of a homogeneous chemical composition of the nebulae. If abundance inhomogeneities are present, one must investigate the astrophysical significance of the abundances measured. Stasi\'nska et al. (2007) have examined this issue in the case of H\,{\sc ii} regions.

\subsection{Shocks and related issues}
   
Photoionization is not the only process leading to the formation of emission lines. Cooling flows (Cox \& Smith 1976, Fabian \& Nulsen 1977)  and interstellar shocks    can also produce them. Shocks are ubiquitous in galaxies, caused by jets  (jets associated to brown dwarfs, protostars or massive black holes), winds (stellar winds,  protostars, active galactic nuclei) or supersonic turbulence. Dopita \& Sutherland (1995, 1996) provide  a grid of models of line emission produced by pure shocks.
However, the efficiency of ionization by stellar photons is such that, whenever hot stars are present, they are likely to dominate the ionization budget of the surrounding gas. The same can probably be said for radiation arising from accretion onto a massive black hole. However, it is true that, in order to explain observed emission line ratios,  the possible contribution of shock-heating to photoionization must be examined.

The effects of shocks that leave a signature on emission line spectra in photoionized nebulae can be summarized as follows:

1) Shocks generate compression and thus high gas densities.

2) This compression locally reduces the ionization parameter, and thus enhances the low ionization emission lines such as [O\,{\sc ii}], [N\,{\sc ii}], [S\,{\sc ii}], and [O\,{\sc i}]. 

3) Shock heating produces very high temperatures (of the order of millions of K), which lead to collisional ionization, and the production of such ions as O$^{3+}$ or He$^{++}$. 

4) This high temperature gas emits X-rays, which contribute to the ionization in front of the shock and behind it.

5) In the region between the X-ray emitting zone and the low excitation zone, the temperature of the gas passes through intermediate temperatures of several tens of thousands of K, boosting the emission of ultraviolet CELs and of auroral lines. The [O\,{\sc iii}] $\lambda$4363/5007 ratio can be significantly enhanced.

However, it must be realized that effects 2, 3 and possibly 5 are spectroscopically indistinguishable from the effects of photoionization by an external X-ray source. As for the enhancement of the low ionization lines, it can simply be due to a small ionization parameter without the necessity of invoking shock heating (see e.g. the explanation proposed by Stasi\'nska \& Izotov 2003 for the behaviour of the  [O\,{\sc i}] lines in H\,{\sc ii} galaxies). 

Finally, dynamical heating and cooling as well as non equilibrium photoionization may be an issue in certain cases, when the dynamical times are shorter than the radiative time. Only a few works have started exploring these avenues, which require important computing power and therefore cannot use the same approach to real nebulae as the ones discussed in the present text. An enlightening introduction to the role of dynamics in photoionized nebulae is given by Henney (2006).

%\appendix

% ------------------------------------------------------------------------------
%
% ------------------------------------------------------------------------------

\section{Appendix: Lists of useful lines and how to deal with them}

In this appendix, we list all the forbidden, semi-forbidden and resonance lines from  C, N, O, Ne, S, Cl and Ar ions which have ionization potentials smaller than 126~eV. These lines as well as all the information on them were extracted from the Atomic Line List compiled by Peter van Hoof and accessible at http://www.pa.uky.edu/\~{}peter/atomic. Permitted lines are far too numerous to be listed there. They are available in the Atomic Line List. However it is more convenient, for a first approach, to use the identifications made by Zhang et al. (2005) on their deep, high resolution spectrum of the very high excitation planetary nebula NGC\,7027. This spectrum covers the entire range between 3300 and 9130 \AA.

Table 1.1 gives 
the electronic configurations and ionization potentials of the C, N, O, Ne, S, Cl and Ar ions. Those ions corresponding to the same isoelectronic sequence have been listed in the same column. Such ions have similar families of lines and provide the same plasma diagnostics. Note that H\,{\sc ii} regions contain only traces (if any) of ions with ionization potential larger than 54~eV. On the other hand, planetary nebulae, which are generally ionized by hotter stars, can have an important fraction of  Ne\,{\sc v} and ions of similar ionization potentials.

\begin{table}[h!]
\begin{center}
\scriptsize{
\begin{tabular}{lllllllll}
\hline
\noalign{\smallskip}
    &    2s    &    2s$^2$    &    2p    &     2p$^2$    &    2p$^3$    &    2p$^4$    &    2p$^5$    &     2p$^6$    \\
\hline
\noalign{\smallskip}
C    &    C\,{\sc iv}    &    C\,{\sc iii}    &     C\,{\sc ii}    &    C\,{\sc i}    &         &         &        &        \\
    &    64.49    &    47.89    &    24.38    &    11.26     &        &        &        &     \\
\noalign{\smallskip}
N    &    N\,{\sc v}    &    N\,{\sc iv}    &     N\,{\sc iii}    &    N\,{\sc ii}    &    N\,{\sc i}     &        &        &        \\
    &    97.89    &    77.47    &    47.45    &    29.60     &    14.53    &        &        &     \\
\noalign{\smallskip}
O    &         &    O\,{\sc v}    &    O\,{\sc iv}     &    O\,{\sc iii}    &    O\,{\sc ii}    &     O\,{\sc i}    &        &        \\
    &        &    113.9    &    77.41    &    54.93     &    35.12    &    13.61    &        &     \\
\noalign{\smallskip}
Ne    &        &        &         &     Ne\,{\sc v}    &    Ne\,{\sc iv}    &    Ne\,{\sc iii}     &    Ne\,{\sc ii}    &    Ne\,{\sc i}    \\
    &        &        &        &     126.21    &    97.11    &    63.45    &    40.46    &     21.56    \\
\hline
\noalign{\smallskip}
    &    3s    &    3s$^2$    &    3p    &     3p$^2$    &    3p$^3$    &    3p$^4$    &    3p$^5$    &     3p$^6$    \\
\hline
\noalign{\smallskip}
S    &    S\,{\sc vi}    &    S\,{\sc v}    &     S\,{\sc iv}    &    S\,{\sc iii}    &    S\,{\sc ii}     &    S\,{\sc i}    &        &        \\
    &    88.05    &    72.68    &    47.30    &    34.83     &    23.33    &    10.36    &        &     \\
\noalign{\smallskip}
Cl    &    Cl\,{\sc vii}    &    Cl~{\sc vi}    &     Cl\,{\sc v}    &    Cl\,{\sc iv}    &    Cl\,{\sc iii}     &    Cl\,{\sc ii}    &    Cl\,{\sc i}    &     \\
    &    114.19    &    93.03    &    67.70    &    53.46     &    39.61    &    23.81    &    12.97    &     \\
\noalign{\smallskip}
Ar    &        &    Ar\,{\sc vii}    &    Ar\,{\sc vi}     &    Ar\,{\sc v}    &    Ar\,{\sc iv}    &     Ar\,{\sc iii}    &    Ar\,{\sc ii}    &    Ar\,{\sc i}     \\
    &        &    124.4    &    91.01    &    75.04     &    59.81    &    40.74    &    27.63    &    15.76     \\
\hline    
\end{tabular}
}
\normalsize
\rm
\caption{ Electronic configurations and ionization potentials (in eV) of ions of C, N, O, Ne, S, Cl, Ar with ionization potentials up to 126~eV.}\label{t1}
\end{center}
\end{table}

% ------------------------------------------------------------------------------
%
%

Tables 1.2--1.8 list, for the ions of C, N, O, Ne, S, Cl, Ar respectively, the  vacuum wavelength of each line (in \AA), the identification, the type of transition (TT), the electronic configurations for the lower and upper levels, the  spectroscopic terms for the lower and upper level (J-J), the transition probability, and the energies of the lower and upper level. 

Note that it is more common to use the wavelength in the air than the wavelength  in the vacuum. The IAU standard for conversion from air to vacuum wavelengths is given in Morton (1991). For vacuum wavelengths   $\lambda_{\rm a}$ in Angstroms, convert to air wavelength   $\lambda_{\rm a} $ via:
  $\lambda_{\rm a} = \lambda_{\rm v} / (1.0 + 2.735182 \times 10^{-4} + 131.4182 / \lambda_{\rm v}^2 + 2.76249\times 10^{8} / \lambda_{\rm v}^4)$.
  
The identification in the tables is represented as:  Fe I for allowed transitions, Fe I] for intercombination transitions, [Fe I] for forbidden transitions.

The symbols for the transition types are:
E1: allowed transitions; M1: magnetic dipole forbidden transitions; E2:
electric quadrupole forbidden transitions.

The transition probabilities have been  taken from the Atomic Line List of Peter van Hoof, or from the Chianti Atomic Database   at \linebreak http://www.arcetri.astro.it/science/chianti/chianti.html if followed by an asterisk.

\begin{table}[h!]
\begin{center}
\scriptsize{
\begin{tabular}{rlcr@{ -- }lr@{ -- }llr@{ -- }l}
\hline
\noalign{\smallskip}
$\lambda_{\rm vac}$ (\AA) \phantom{00}&Ion&TT&\multicolumn{2}{c}{Terms}&$J$&$J$&\multicolumn{1}{c}{$A_{\rm ki}$ (s$^{-1}$)}&\multicolumn{2}{c}{Levels (cm$^{-1}$)}\\
\hline
\noalign{\smallskip}
   4622.864\phantom{000}	 & 	\llap{[\,}C\,{\sc i}\,]	 & 	M1	 & 	$^3\!P$   	 & 	$^1\!S$   	 & 	$1$ 	 & 	$0$ 	 & 	$$      	 & 	       16.40 	 & 	    21648.01\\
   8729.52\phantom{0000}	 & 	\llap{[\,}C\,{\sc i}\,]	 & 	E2	 & 	$^1\!D$   	 & 	$^1\!S$   	 & 	$2$ 	 & 	$0$ 	 & 	$$      	 & 	    10192.63 	 & 	    21648.01\\
   9811.01\phantom{0000}	 & 	\llap{[\,}C\,{\sc i}\,]	 & 	E2	 & 	$^3\!P$   	 & 	$^1\!D$   	 & 	$0$ 	 & 	$2$ 	 & 	$$      	 & 	        0.00 	 & 	    10192.63\\
   9826.82\phantom{0000}	 & 	\llap{[\,}C\,{\sc i}\,]	 & 	M1	 & 	$^3\!P$   	 & 	$^1\!D$   	 & 	$1$ 	 & 	$2$ 	 & 	$$      	 & 	       16.40 	 & 	    10192.63\\
   9852.96\phantom{0000}	 & 	\llap{[\,}C\,{\sc i}\,]	 & 	M1	 & 	$^3\!P$   	 & 	$^1\!D$   	 & 	$2$ 	 & 	$2$ 	 & 	$$      	 & 	       43.40 	 & 	    10192.63\\
2304000.\phantom{000000}	 & 	\llap{[\,}C\,{\sc i}\,]	 & 	E2	 & 	$^3\!P$   	 & 	$^3\!P$   	 & 	$0$ 	 & 	$2$ 	 & 	$$      	 & 	        0.00 	 & 	       43.40\\
3704000.\phantom{000000}	 & 	\llap{[\,}C\,{\sc i}\,]	 & 	M1	 & 	$^3\!P$   	 & 	$^3\!P$   	 & 	$1$ 	 & 	$2$ 	 & 	$2.65(-7)$	 & 	       16.40 	 & 	       43.40\\
6100000.\phantom{000000}	 & 	\llap{[\,}C\,{\sc i}\,]	 & 	M1	 & 	$^3\!P$   	 & 	$^3\!P$   	 & 	$0$ 	 & 	$1$ 	 & 	$7.93(-8)$	 & 	        0.00 	 & 	       16.40\\
\hline
\noalign{\smallskip}
   1334.5323\phantom{00} & C\,{\sc ii}               & E1 & $^{2}{\rm P}^\circ$ & $^{2}{\rm D}$       & $\frac{1}{2}$ & $\frac{3}{2}$ & $2.38(+8)$ &  $0.00$ & $74932.62$ \\
   1335.6627\phantom{00} & C\,{\sc ii}               & E1 & $^{2}{\rm P}^\circ$ & $^{2}{\rm D}$       & $\frac{3}{2}$ & $\frac{3}{2}$ & $4.74(+7)$ & $63.42$ & $74932.62$ \\
   1335.7077\phantom{00} & C\,{\sc ii}               & E1 & $^{2}{\rm P}^\circ$ & $^{2}{\rm D}$       & $\frac{3}{2}$ & $\frac{5}{2}$ & $2.84(+8)$ & $63.42$ & $74930.10$ \\
   2322.69\phantom{0000} & \llap{[\,}C\,{\sc ii}\,]  & M2 & $^{2}{\rm P}^\circ$ & $^{4}{\rm P}$       & $\frac{1}{2}$ & $\frac{5}{2}$ &          &  $0.00$ & $43053.60$ \\
   2324.21\phantom{0000} & C\,{\sc ii}\,]            & E1 & $^{2}{\rm P}^\circ$ & $^{4}{\rm P}$       & $\frac{1}{2}$ & $\frac{3}{2}$ &          &  $0.00$ & $43025.30$ \\
   2325.40\phantom{0000} & C\,{\sc ii}\,]            & E1 & $^{2}{\rm P}^\circ$ & $^{4}{\rm P}$       & $\frac{1}{2}$ & $\frac{1}{2}$ &          &  $0.00$ & $43003.30$ \\
   2326.11\phantom{0000} & C\,{\sc ii}\,]            & E1 & $^{2}{\rm P}^\circ$ & $^{4}{\rm P}$       & $\frac{3}{2}$ & $\frac{5}{2}$ &          & $63.42$ & $43053.60$ \\
   2327.64\phantom{0000} & C\,{\sc ii}\,]            & E1 & $^{2}{\rm P}^\circ$ & $^{4}{\rm P}$       & $\frac{3}{2}$ & $\frac{3}{2}$ &          & $63.42$ & $43025.30$ \\
   2328.84\phantom{0000} & C\,{\sc ii}\,]            & E1 & $^{2}{\rm P}^\circ$ & $^{4}{\rm P}$       & $\frac{3}{2}$ & $\frac{1}{2}$ &          & $63.42$ & $43003.30$ \\
1576800.\phantom{000000} & 	\llap{[\,}C\,{\sc ii}\,] & M1 & $^2\!P^\circ$	    & 	$^2\!P^\circ$	  &	$\frac{ 1}{2}$ & $\frac{ 3}{2} $	 & 	$2.290(-6)^*$	 & 	        0.00 	 & 	       63.42\\
4550000.\phantom{000000} & \llap{[\,}C\,{\sc ii}\,]  & M1 &       $^{4}{\rm P}$ & $^{4}{\rm P}$       & $\frac{1}{2}$ & $\frac{3}{2}$ & $2.39(-7)$ & $43003.30$ & $43025.30$ \\
\noalign{\smallskip}
\hline
\noalign{\smallskip}
   1906.683\phantom{000}   & \llap{[\,}C\,{\sc iii}\,] & M2 &       $^{1}{\rm S}$ & $^{3}{\rm P}^\circ$ &           $0$ & $2$           &          &  $0.00$ & $52447.11$ \\
   1908.734\phantom{000}   & C\,{\sc iii}\,]           & E1 &       $^{1}{\rm S}$ & $^{3}{\rm P}^\circ$ &           $0$ & $1$           &          &  $0.00$ & $52390.75$ \\
1249200.\phantom{000000}   & 	\llap{[\,}C\,{\sc iii}\,]	 & 	E2	 & 	$^3\!P^\circ$	 & 	$^3\!P^\circ$	 & 	$0$ 	 & 	$2$ 	 & 	$$      	 & 	    52367.06 	 & 	    52447.11\\
1774300.\phantom{000000}	 & 	\llap{[\,}C\,{\sc iii}\,]	 & 	M1	 & 	$^3\!P^\circ$	 & 	$^3\!P^\circ$	 & 	$1$ 	 & 	$2$ 	 & 	$2.450(-6)^*$	 & 	    52390.75 	 & 	    52447.11\\
4221000.\phantom{000000}	 & 	\llap{[\,}C\,{\sc iii}\,]	 & 	M1	 & 	$^3\!P^\circ$	 & 	$^3\!P^\circ$	 & 	$0$ 	 & 	$1$ 	 & 	$2.39(-7)$	 & 	    52367.06 	 & 	    52390.75\\
\hline
\noalign{\smallskip}
   1548.203\phantom{000}   & C\,{\sc iv}               & E1 &       $^{2}{\rm S}$ & $^{2}{\rm P}^\circ$ & $\frac{1}{2}$ & $\frac{3}{2}$ & $2.65(+8)$ &  $0.00$ & $64591.00$ \\
   1550.777\phantom{000}   & C\,{\sc iv}               & E1 &       $^{2}{\rm S}$ & $^{2}{\rm P}^\circ$ & $\frac{1}{2}$ & $\frac{1}{2}$ & $2.64(+8)$ &  $0.00$ & $64483.80$ \\
\hline
\end{tabular}
}
\normalsize
\rm
\caption{ Lines of C ions. }\label{t1}
\end{center}
\end{table}

\newpage
																		
% ------------------------------------------------------------------------------
%
% ------------------------------------------------------------------------------

\begin{table}[h!]
\begin{center}
\scriptsize{
\begin{tabular}{rlcr@{ -- }lr@{ -- }llr@{ -- }l}
\hline
\noalign{\smallskip}
$\lambda_{\rm vac}$ (\AA) \phantom{00}&Ion&TT&\multicolumn{2}{c}{Terms}&$J$&$J$&\multicolumn{1}{c}{$A_{\rm ki}$ (s$^{-1}$)}&\multicolumn{2}{c}{Levels (cm$^{-1}$)}\\
\hline
\noalign{\smallskip}
   3467.4898\phantom{00}	 & 	\llap{[\,}N\,{\sc i}\,]	 & 	M1	 & 	$^4\!S^\circ$	 & 	$^2\!P^\circ$	 & 	$\frac{ 3}{2}$	 & 	$\frac{ 3}{2}$	 & 	$6.210(-3)^*$      	 & 	        0.00 	 & 	    28839.31\\
   3467.5362\phantom{00}	 & 	\llap{[\,}N\,{\sc i}\,]	 & 	M1	 & 	$^4\!S^\circ$	 & 	$^2\!P^\circ$	 & 	$\frac{ 3}{2}$	 & 	$\frac{ 1}{2}$	 & 	$2.520(-3)^*$      	 & 	        0.00 	 & 	    28838.92\\
   5199.3490\phantom{00}	 & 	\llap{[\,}N\,{\sc i}\,]	 & 	E2	 & 	$^4\!S^\circ$	 & 	$^2\!D^\circ$	 & 	$\frac{ 3}{2}$	 & 	$\frac{ 3}{2}$	 & 	$2.260(-5)^*$      	 & 	        0.00 	 & 	    19233.18\\
   5201.7055\phantom{00}	 & 	\llap{[\,}N\,{\sc i}\,]	 & 	E2	 & 	$^4\!S^\circ$	 & 	$^2\!D^\circ$	 & 	$\frac{ 3}{2}$	 & 	$\frac{ 5}{2}$	 & 	$5.765(-6)^*$      	 & 	        0.00 	 & 	    19224.46\\
  10400.587\phantom{000}	 & 	\llap{[\,}N\,{\sc i}\,]	 & 	M1	 & 	$^2\!D^\circ$	 & 	$^2\!P^\circ$	 & 	$\frac{ 5}{2}$	 & 	$\frac{ 3}{2}$	 & 	$5.388(-2)^*$      	 & 	    19224.46 	 & 	    28839.31\\
  10401.004\phantom{000}	 & 	\llap{[\,}N\,{\sc i}\,]	 & 	E2	 & 	$^2\!D^\circ$	 & 	$^2\!P^\circ$	 & 	$\frac{ 5}{2}$	 & 	$\frac{ 1}{2}$	 & 	$3.030(-2)^*$      	 & 	    19224.46 	 & 	    28838.92\\
  10410.021\phantom{000}	 & 	\llap{[\,}N\,{\sc i}\,]	 & 	M1	 & 	$^2\!D^\circ$	 & 	$^2\!P^\circ$	 & 	$\frac{ 3}{2}$	 & 	$\frac{ 3}{2}$	 & 	$2.435(-2)^*$      	 & 	    19233.18 	 & 	    28839.31\\
  10410.439\phantom{000}	 & 	\llap{[\,}N\,{\sc i}\,]	 & 	M1	 & 	$^2\!D^\circ$	 & 	$^2\!P^\circ$	 & 	$\frac{ 3}{2}$	 & 	$\frac{ 1}{2}$	 & 	$4.629(-2)^*$      	 & 	    19233.18 	 & 	    28838.92\\
\hline
\noalign{\smallskip}
   2137.457\phantom{000} & \llap{[\,}N\,{\sc ii}\,]  & M2 &       $^{3}{\rm P}$ & $^{5}{\rm S}^\circ$ &           $0$ & $2$           &          &   $0.00$ & $46784.56$ \\
   2139.683\phantom{000} & N\,{\sc ii}\,]            & E1 &       $^{3}{\rm P}$ & $^{5}{\rm S}^\circ$ &           $1$ & $2$           &          &  $48.67$ & $46784.56$ \\
   2143.450\phantom{000} & N\,{\sc ii}\,]            & E1 &       $^{3}{\rm P}$ & $^{5}{\rm S}^\circ$ &           $2$ & $2$           &          & $130.80$ & $46784.56$ \\
   3063.728\phantom{000}	 & 	\llap{[\,}N\,{\sc ii}\,]	 & 	M1	 & 	$^3\!P$   	 & 	$^1\!S$   	 & 	$1$ 	 & 	$0$ 	 & 	$4.588(-6)^*$      	 & 	       48.67 	 & 	    32688.64\\
   3071.457\phantom{000} & \llap{[\,}N\,{\sc ii}\,]  & E2 &       $^{3}{\rm P}$ & $^{1}{\rm S}$       &           $2$ & $0$           &          &   $130.80$ & $32688.64$ \\
   5756.240\phantom{000}	 & 	\llap{[\,}N\,{\sc ii}\,]	 & 	E2	 & 	$^1\!D$   	 & 	$^1\!S$   	 & 	$2$ 	 & 	$0$ 	 & 	$9.234(-1)^*$      	 & 	    15316.19 	 & 	    32688.64\\
   6529.04\phantom{0000}	 & 	\llap{[\,}N\,{\sc ii}\,]	 & 	E2	 & 	$^3\!P$   	 & 	$^1\!D$   	 & 	$0$ 	 & 	$2$ 	 & 	$1.928(-6)^*$      	 & 	        0.00 	 & 	    15316.19\\
   6549.85\phantom{0000}	 & 	\llap{[\,}N\,{\sc ii}\,]	 & 	M1	 & 	$^3\!P$   	 & 	$^1\!D$   	 & 	$1$ 	 & 	$2$ 	 & 	$9.819(-4)^*$      	 & 	       48.67 	 & 	    15316.19\\
   6585.28\phantom{0000}	 & 	\llap{[\,}N\,{\sc ii}\,]	 & 	M1	 & 	$^3\!P$   	 & 	$^1\!D$   	 & 	$2$ 	 & 	$2$ 	 & 	$3.015(-3)^*$      	 & 	      130.80 	 & 	    15316.19\\
 764500.\phantom{000000}	 & 	\llap{[\,}N\,{\sc ii}\,]	 & 	E2	 & 	$^3\!P$   	 & 	$^3\!P$   	 & 	$0$ 	 & 	$2$ 	 & 	$$      	 & 	        0.00 	 & 	      130.80\\
1217600.\phantom{000000}	 & 	\llap{[\,}N\,{\sc ii}\,]	 & 	M1	 & 	$^3\!P$   	 & 	$^3\!P$   	 & 	$1$ 	 & 	$2$ 	 & 	$7.460(-6)^*$	 & 	       48.67 	 & 	      130.80\\
2055000.\phantom{000000}	 & 	\llap{[\,}N\,{\sc ii}\,]	 & 	M1	 & 	$^3\!P$   	 & 	$^3\!P$   	 & 	$0$ 	 & 	$1$ 	 & 	$2.080(-6)^*$	 & 	        0.00 	 & 	       48.67\\
\hline
\noalign{\smallskip}
   1744.351\phantom{000} & \llap{[\,}N\,{\sc iii}\,] & M2 & $^{2}{\rm P}^\circ$ & $^{4}{\rm P}$       & $\frac{1}{2}$ & $\frac{5}{2}$ &          &   $0.00$ & $57327.90$ \\
   1746.823\phantom{000} & N\,{\sc iii}\,]           & E1 & $^{2}{\rm P}^\circ$ & $^{4}{\rm P}$       & $\frac{1}{2}$ & $\frac{3}{2}$ &          &   $0.00$ & $57246.80$ \\
   1748.646\phantom{000} & N\,{\sc iii}\,]           & E1 & $^{2}{\rm P}^\circ$ & $^{4}{\rm P}$       & $\frac{1}{2}$ & $\frac{1}{2}$ &          &   $0.00$ & $57187.10$ \\
   1749.674\phantom{000} & N\,{\sc iii}\,]           & E1 & $^{2}{\rm P}^\circ$ & $^{4}{\rm P}$       & $\frac{3}{2}$ & $\frac{5}{2}$ &          & $174.40$ & $57327.90$ \\
   1752.160\phantom{000} & N\,{\sc iii}\,]           & E1 & $^{2}{\rm P}^\circ$ & $^{4}{\rm P}$       & $\frac{3}{2}$ & $\frac{3}{2}$ &          & $174.40$ & $57246.80$ \\
   1753.995\phantom{000} & N\,{\sc iii}\,]           & E1 & $^{2}{\rm P}^\circ$ & $^{4}{\rm P}$       & $\frac{3}{2}$ & $\frac{1}{2}$ &          & $174.40$ & $57187.10$ \\
 573400.\phantom{000000}	 & 	\llap{[\,}N\,{\sc iii}\,]	 & 	M1	 & 	$^2\!P^\circ$	 & 	$^2\!P^\circ$	 & 	$\frac{ 1}{2}$	 & 	$\frac{ 3}{2}$	 & 	$4.770(-5)^*$	 & 	        0.00 	 & 	      174.40\\
 710000.\phantom{000000} & \llap{[\,}N\,{\sc iii}\,] & E2 &       $^{4}{\rm P}$ & $^{4}{\rm P}$       & $\frac{1}{2}$ & $\frac{5}{2}$ &          & $57187.10$ & $57327.90$ \\
1233000.\phantom{000000} & \llap{[\,}N\,{\sc iii}\,] & M1 &       $^{4}{\rm P}$ & $^{4}{\rm P}$       & $\frac{3}{2}$ & $\frac{5}{2}$ & $8.63(-6)$ & $57246.80$ & $57327.90$ \\
1675000.\phantom{000000} & \llap{[\,}N\,{\sc iii}\,] & M1 &       $^{4}{\rm P}$ & $^{4}{\rm P}$       & $\frac{1}{2}$ & $\frac{3}{2}$ & $4.78(-6)$ & $57187.10$ & $57246.80$ \\
\hline
\noalign{\smallskip}
   1483.321\phantom{000} & \llap{[\,}N\,{\sc iv}\,]  & M2 &       $^{1}{\rm S}$ & $^{3}{\rm P}^\circ$ &           $0$ & $2$           &          &   $0.00$ & $67416.30$ \\
   1486.496\phantom{000} & N\,{\sc iv}\,]            & E1 &       $^{1}{\rm S}$ & $^{3}{\rm P}^\circ$ &           $0$ & $1$           &          &   $0.00$ & $67272.30$ \\
 482900.\phantom{000000}	 & 	\llap{[\,}N\,{\sc iv}\,]	 & 	E2	 & 	$^3\!P^\circ$	 & 	$^3\!P^\circ$	 & 	$0$ 	 & 	$2$ 	 & 	$$      	 & 	    67209.20 	 & 	    67416.30\\
 694000.\phantom{000000}	 & 	\llap{[\,}N\,{\sc iv}\,]	 & 	M1	 & 	$^3\!P^\circ$	 & 	$^3\!P^\circ$	 & 	$1$ 	 & 	$2$ 	 & 	$4.027(-5)^*$	 & 	    67272.30 	 & 	    67416.30\\
1585000.\phantom{000000}	 & 	\llap{[\,}N\,{\sc iv}\,]	 & 	M1	 & 	$^3\!P^\circ$	 & 	$^3\!P^\circ$	 & 	$0$ 	 & 	$1$ 	 & 	$4.52(-6)$	 & 	    67209.20 	 & 	    67272.30\\
\hline
\end{tabular}
}
\normalsize
\rm
\caption{ Lines of N ions.}\label{t1}
\end{center}
\end{table}

\newpage

% ------------------------------------------------------------------------------
%
% ------------------------------------------------------------------------------

\begin{table}[h!]
\begin{center}
\scriptsize{
\begin{tabular}{rlcr@{ -- }lr@{ -- }llr@{ -- }l}
\hline
\noalign{\smallskip}
$\lambda_{\rm vac}$ (\AA) \phantom{00}&Ion&TT&\multicolumn{2}{c}{Terms}&$J$&$J$&\multicolumn{1}{c}{$A_{\rm ki}$ (s$^{-1}$)}&\multicolumn{2}{c}{Levels (cm$^{-1}$)}\\
\hline
\noalign{\smallskip}
   2973.1538\phantom{00}	 & 	\llap{[\,}O\,{\sc i}\,]	 & 	M1	 & 	$^3\!P$   	 & 	$^1\!S$   	 & 	$1$ 	 & 	$0$ 	 & 	$$      	 & 	      158.26 	 & 	    33792.58\\
   5578.8874\phantom{00}	 & 	\llap{[\,}O\,{\sc i}\,]	 & 	E2	 & 	$^1\!D$   	 & 	$^1\!S$   	 & 	$2$ 	 & 	$0$ 	 & 	$$      	 & 	    15867.86 	 & 	    33792.58\\
   6302.046\phantom{000}	 & 	\llap{[\,}O\,{\sc i}\,]	 & 	M1	 & 	$^3\!P$   	 & 	$^1\!D$   	 & 	$2$ 	 & 	$2$ 	 & 	$$      	 & 	        0.00 	 & 	    15867.86\\
   6365.536\phantom{000}	 & 	\llap{[\,}O\,{\sc i}\,]	 & 	M1	 & 	$^3\!P$   	 & 	$^1\!D$   	 & 	$1$ 	 & 	$2$ 	 & 	$$      	 & 	      158.26 	 & 	    15867.86\\
   6393.500\phantom{000}	 & 	\llap{[\,}O\,{\sc i}\,]	 & 	E2	 & 	$^3\!P$   	 & 	$^1\!D$   	 & 	$0$ 	 & 	$2$ 	 & 	$$      	 & 	      226.98 	 & 	    15867.86\\
 440573.\phantom{000000}	 & 	\llap{[\,}O\,{\sc i}\,]	 & 	E2	 & 	$^3\!P$   	 & 	$^3\!P$   	 & 	$2$ 	 & 	$0$ 	 & 	$$      	 & 	        0.00 	 & 	      226.98\\
 631850.\phantom{000000}	 & 	\llap{[\,}O\,{\sc i}\,]	 & 	M1	 & 	$^3\!P$   	 & 	$^3\!P$   	 & 	$2$ 	 & 	$1$ 	 & 	$8.91(-5)$	 & 	        0.00 	 & 	      158.26\\
1455350.\phantom{000000}	 & 	\llap{[\,}O\,{\sc i}\,]	 & 	M1	 & 	$^3\!P$   	 & 	$^3\!P$   	 & 	$1$ 	 & 	$0$ 	 & 	$1.75(-5)$	 & 	      158.26 	 & 	      226.98\\
\hline
\noalign{\smallskip}
   2470.966\phantom{000}	 & 	\llap{[\,}O\,{\sc ii}\,]	 & 	M1	 & 	$^4\!S^\circ$	 & 	$^2\!P^\circ$	 & 	$\frac{ 3}{2}$	 & 	$\frac{ 1}{2}$	 & 	$2.380(-2)^*$      	 & 	        0.00 	 & 	    40470.00\\
   2471.088\phantom{000}	 & 	\llap{[\,}O\,{\sc ii}\,]	 & 	M1	 & 	$^4\!S^\circ$	 & 	$^2\!P^\circ$	 & 	$\frac{ 3}{2}$	 & 	$\frac{ 3}{2}$	 & 	$5.800(-2)^*$      	 & 	        0.00 	 & 	    40468.01\\
   3727.092\phantom{000}	 & 	\llap{[\,}O\,{\sc ii}\,]	 & 	E2	 & 	$^4\!S^\circ$	 & 	$^2\!D^\circ$	 & 	$\frac{ 3}{2}$	 & 	$\frac{ 3}{2}$	 & 	$1.810(-4)^*$      	 & 	        0.00 	 & 	    26830.57\\
   3729.875\phantom{000}	 & 	\llap{[\,}O\,{\sc ii}\,]	 & 	E2	 & 	$^4\!S^\circ$	 & 	$^2\!D^\circ$	 & 	$\frac{ 3}{2}$	 & 	$\frac{ 5}{2}$	 & 	$3.588(-5)^*$      	 & 	        0.00 	 & 	    26810.55\\
   7320.94\phantom{0000}	 & 	\llap{[\,}O\,{\sc ii}\,]	 & 	E2	 & 	$^2\!D^\circ$	 & 	$^2\!P^\circ$	 & 	$\frac{ 5}{2}$	 & 	$\frac{ 1}{2}$	 & 	$5.630(-2)^*$      	 & 	    26810.55 	 & 	    40470.00\\
   7322.01\phantom{0000}	 & 	\llap{[\,}O\,{\sc ii}\,]	 & 	M1	 & 	$^2\!D^\circ$	 & 	$^2\!P^\circ$	 & 	$\frac{ 5}{2}$	 & 	$\frac{ 3}{2}$	 & 	$1.074(-1)^*$      	 & 	    26810.55 	 & 	    40468.01\\
   7331.68\phantom{0000}	 & 	\llap{[\,}O\,{\sc ii}\,]	 & 	M1	 & 	$^2\!D^\circ$	 & 	$^2\!P^\circ$	 & 	$\frac{ 3}{2}$	 & 	$\frac{ 1}{2}$	 & 	$9.410(-2)^*$      	 & 	    26830.57 	 & 	    40470.00\\
   7332.75\phantom{0000}	 & 	\llap{[\,}O\,{\sc ii}\,]	 & 	M1	 & 	$^2\!D^\circ$	 & 	$^2\!P^\circ$	 & 	$\frac{ 3}{2}$	 & 	$\frac{ 3}{2}$	 & 	$5.800(-2)^*$      	 & 	    26830.57 	 & 	    40468.01\\
4995000.\phantom{000000}	 & 	\llap{[\,}O\,{\sc ii}\,]	 & 	M1	 & 	$^2\!D^\circ$	 & 	$^2\!D^\circ$	 & 	$\frac{ 5}{2}$	 & 	$\frac{ 3}{2}$	 & 	$1.320(-7)^*$         & 	    26810.55 	 & 	    26830.57\\
\hline
\noalign{\smallskip}
   1657.6933\phantom{00}     & \llap{[\,}O\,{\sc iii}\,] & M2 &       $^{3}{\rm P}$ & $^{5}{\rm S}^\circ$ &           $0$ & $2$           &          &   $0.00$ & $60324.79$ \\
   1660.8092\phantom{00}     & O\,{\sc iii}\,]           & E1 &       $^{3}{\rm P}$ & $^{5}{\rm S}^\circ$ &           $1$ & $2$           &          & $113.18$ & $60324.79$ \\
   1666.1497\phantom{00}     & O\,{\sc iii}\,]           & E1 &       $^{3}{\rm P}$ & $^{5}{\rm S}^\circ$ &           $2$ & $2$           &          & $306.17$ & $60324.79$ \\
   2321.664\phantom{000}     & \llap{[\,}O\,{\sc iii}\,] & M1 &       $^{3}{\rm P}$ & $^{1}{\rm S}$       &           $1$ & $0$           &          & $113.18$ & $43185.74$ \\
   2332.113\phantom{000}     & \llap{[\,}O\,{\sc iii}\,] & E2 &       $^{3}{\rm P}$ & $^{1}{\rm S}$       &           $2$ & $0$           &          & $306.17$ & $43185.74$ \\
   4364.436\phantom{000}	 & 	\llap{[\,}O\,{\sc iii}\,]	 & 	E2	 & 	$^1\!D$   	 & 	$^1\!S$   	 & 	$2$ 	 & 	$0$ 	 & 	$4.760(-1)^*$      	 & 	    20273.27 	 & 	    43185.74\\
   4932.603\phantom{000}	 & 	\llap{[\,}O\,{\sc iii}\,]	 & 	E2	 & 	$^3\!P$   	 & 	$^1\!D$   	 & 	$0$ 	 & 	$2$ 	 & 	$7.250(-6)^*$      	 & 	        0.00 	 & 	    20273.27\\
   4960.295\phantom{000}	 & 	\llap{[\,}O\,{\sc iii}\,]	 & 	M1	 & 	$^3\!P$   	 & 	$^1\!D$   	 & 	$1$ 	 & 	$2$ 	 & 	$6.791(-3)^*$      	 & 	      113.18 	 & 	    20273.27\\
   5008.240\phantom{000}	 & 	\llap{[\,}O\,{\sc iii}\,]	 & 	M1	 & 	$^3\!P$   	 & 	$^1\!D$   	 & 	$2$ 	 & 	$2$ 	 & 	$2.046(-2)^*$      	 & 	      306.17 	 & 	    20273.27\\
 326612.\phantom{000000}	 & 	\llap{[\,}O\,{\sc iii}\,]	 & 	E2	 & 	$^3\!P$   	 & 	$^3\!P$   	 & 	$0$ 	 & 	$2$ 	 & 	$$      	 & 	        0.00 	 & 	      306.17\\
 518145.\phantom{000000}	 & 	\llap{[\,}O\,{\sc iii}\,]	 & 	M1	 & 	$^3\!P$   	 & 	$^3\!P$   	 & 	$1$ 	 & 	$2$ 	 & 	$1.010(-4)^*$	 & 	      113.18 	 & 	      306.17\\
 883560.\phantom{000000}	 & 	\llap{[\,}O\,{\sc iii}\,]	 & 	M1	 & 	$^3\!P$   	 & 	$^3\!P$   	 & 	$0$ 	 & 	$1$ 	 & 	$2.700(-5)^*$	 & 	        0.00 	 & 	      113.18\\
\hline
\noalign{\smallskip}
   1393.621\phantom{000} & \llap{[\,}O\,{\sc iv}\,]  & M2 & $^{2}{\rm P}^\circ$ & $^{4}{\rm P}$       & $\frac{1}{2}$ & $\frac{5}{2}$ &          &   $0.00$ & $71755.50$ \\
   1397.232\phantom{000} & O\,{\sc iv}\,]            & E1 & $^{2}{\rm P}^\circ$ & $^{4}{\rm P}$       & $\frac{1}{2}$ & $\frac{3}{2}$ &          &   $0.00$ & $71570.10$ \\
   1399.780\phantom{000} & O\,{\sc iv}\,]            & E1 & $^{2}{\rm P}^\circ$ & $^{4}{\rm P}$       & $\frac{1}{2}$ & $\frac{1}{2}$ &          &   $0.00$ & $71439.80$ \\
   1401.164\phantom{000} & O\,{\sc iv}\,]            & E1 & $^{2}{\rm P}^\circ$ & $^{4}{\rm P}$       & $\frac{3}{2}$ & $\frac{5}{2}$ &          & $386.25$ & $71755.50$ \\
   1404.813\phantom{000} & O\,{\sc iv}\,]            & E1 & $^{2}{\rm P}^\circ$ & $^{4}{\rm P}$       & $\frac{3}{2}$ & $\frac{3}{2}$ &          & $386.25$ & $71570.10$ \\
   1407.389\phantom{000} & O\,{\sc iv}\,]            & E1 & $^{2}{\rm P}^\circ$ & $^{4}{\rm P}$       & $\frac{3}{2}$ & $\frac{1}{2}$ &          & $386.25$ & $71439.80$ \\
 258903.\phantom{000000}	 & 	\llap{[\,}O\,{\sc iv}\,]	 & 	M1	 & 	$^2\!P^\circ$	 & 	$^2\!P^\circ$	 & 	$\frac{ 1}{2}$	 & 	$\frac{ 3}{2}$	 & 	$5.310(-4)^*$	 & 	        0.00 	 & 	      386.25\\
 316800.\phantom{000000} & \llap{[\,}O\,{\sc iv}\,]  & E2 &       $^{4}{\rm P}$ & $^{4}{\rm P}$       & $\frac{1}{2}$ & $\frac{5}{2}$ &          & $71439.80$ & $71755.50$ \\
 539400.\phantom{000000} & \llap{[\,}O\,{\sc iv}\,]  & M1 &       $^{4}{\rm P}$ & $^{4}{\rm P}$       & $\frac{3}{2}$ & $\frac{5}{2}$ & $1.03(-4)$ & $71570.10$ & $71755.50$ \\
 767000.\phantom{000000} & \llap{[\,}O\,{\sc iv}\,]  & M1 &       $^{4}{\rm P}$ & $^{4}{\rm P}$       & $\frac{1}{2}$ & $\frac{3}{2}$ & $4.97(-5)$ & $71439.80$ & $71570.10$ \\
 \hline
\noalign{\smallskip}
   1213.809\phantom{000}     & \llap{[\,}O\,{\sc v}\,]   & M2 &       $^{1}{\rm S}$ & $^{3}{\rm P}^\circ$ &           $0$ & $2$           &          &     $0.00$ & $82385.30$ \\
 225800.\phantom{000000}	 & 	\llap{[\,}O\,{\sc v}\,]	 & 	E2	 & 	$^3\!P^\circ$	 & 	$^3\!P^\circ$	 & 	$0$ 	 & 	$2$ 	 & 	$$      	 & 	    81942.50 	 & 	    82385.30\\
 326100.\phantom{000000}	 & 	\llap{[\,}O\,{\sc v}\,]	 & 	M1	 & 	$^3\!P^\circ$	 & 	$^3\!P^\circ$	 & 	$1$ 	 & 	$2$ 	 & 	$3.913(-4)^*$	 & 	    82078.60 	 & 	    82385.30\\
 735000.\phantom{000000}	 & 	\llap{[\,}O\,{\sc v}\,]	 & 	M1	 & 	$^3\!P^\circ$	 & 	$^3\!P^\circ$	 & 	$0$ 	 & 	$1$ 	 & 	$4.53(-5)$	 & 	    81942.50 	 & 	    82078.60\\
\hline
\end{tabular}
}
\normalsize
\rm
\caption{ Lines of O ions.}\label{t1}
\end{center}
\end{table}
																		
\newpage

% ------------------------------------------------------------------------------
%
% ------------------------------------------------------------------------------

\begin{table}[h!]
\begin{center}
\scriptsize{
\begin{tabular}{rlcr@{ -- }lr@{ -- }llr@{ -- }l}
\hline
\noalign{\smallskip}
$\lambda_{\rm vac}$ (\AA) \phantom{00}&Ion&TT&\multicolumn{2}{c}{Terms}&$J$&$J$&\multicolumn{1}{c}{$A_{\rm ki}$ (s$^{-1}$)}&\multicolumn{2}{c}{Levels (cm$^{-1}$)}\\
\hline
\noalign{\smallskip}
 128135.48\phantom{0000}	 & 	\llap{[\,}Ne\,{\sc ii}\,]	 & 	M1	 & 	$^2\!P^\circ$	 & 	$^2\!P^\circ$	 & 	$\frac{ 3}{2}$	 & 	$\frac{ 1}{2}$	 & 	$8.590(-3)^*$	 & 	        0.00 	 & 	      780.42\\
\hline
\noalign{\smallskip}
   1793.802\phantom{000} & \llap{[\,}Ne\,{\sc iii}\,] & E2 &       $^{3}{\rm P}$ & $^{1}{\rm S}$       &           $2$ & $0$           &          &    $0.00$ & $55747.50$ \\
   1814.730\phantom{000} & \llap{[\,}Ne\,{\sc iii}\,] & M1 &       $^{3}{\rm P}$ & $^{1}{\rm S}$       &           $1$ & $0$           &          &  $642.88$ & $55747.50$ \\
   3343.50\phantom{0000}	 & 	\llap{[\,}Ne\,{\sc iii}\,]	 & 	E2	 & 	$^1\!D$   	 & 	$^1\!S$   	 & 	$2$ 	 & 	$0$ 	 & 	$2.550$      	 & 	    25838.70 	 & 	    55747.50\\
   3870.16\phantom{0000}	 & 	\llap{[\,}Ne\,{\sc iii}\,]	 & 	M1	 & 	$^3\!P$   	 & 	$^1\!D$   	 & 	$2$ 	 & 	$2$ 	 & 	$1.653(-1)^*$      	 & 	        0.00 	 & 	    25838.70\\
   3968.91\phantom{0000}	 & 	\llap{[\,}Ne\,{\sc iii}\,]	 & 	M1	 & 	$^3\!P$   	 & 	$^1\!D$   	 & 	$1$ 	 & 	$2$ 	 & 	$5.107(-2)^*$      	 & 	      642.88 	 & 	    25838.70\\
   4013.14\phantom{0000}	 & 	\llap{[\,}Ne\,{\sc iii}\,]	 & 	E2	 & 	$^3\!P$   	 & 	$^1\!D$   	 & 	$0$ 	 & 	$2$ 	 & 	$1.555(-5)^*$      	 & 	      920.55 	 & 	    25838.70\\
 108630.5\phantom{00000}	 & 	\llap{[\,}Ne\,{\sc iii}\,]	 & 	E2	 & 	$^3\!P$   	 & 	$^3\!P$   	 & 	$2$ 	 & 	$0$ 	 & 	$2.524(-8)^*$      	 & 	        0.00 	 & 	      920.55\\
 155550.5\phantom{00000}	 & 	\llap{[\,}Ne\,{\sc iii}\,]	 & 	M1	 & 	$^3\!P$   	 & 	$^3\!P$   	 & 	$2$ 	 & 	$1$ 	 & 	$5.971(-3)^*$	 & 	        0.00 	 & 	      642.88\\
 360135.\phantom{000000}	 & 	\llap{[\,}Ne\,{\sc iii}\,]	 & 	M1	 & 	$^3\!P$   	 & 	$^3\!P$   	 & 	$1$ 	 & 	$0$ 	 & 	$1.154(-3)^*$	 & 	      642.88 	 & 	      920.55\\
\hline
\noalign{\smallskip}
   4713.2\phantom{00000}	 & 	\llap{[\,}Ne\,{\sc iv}\,]	 & 	M1	 & 	$^2\!D^\circ$	 & 	$^2\!P^\circ$	 & 	$\frac{ 5}{2}$	 & 	$\frac{ 3}{2}$	 & 	$4.857(-1)^*$      	 & 	    41210.   	 & 	    62427.  \\
   4715.4\phantom{00000}	 & 	\llap{[\,}Ne\,{\sc iv}\,]	 & 	E2	 & 	$^2\!D^\circ$	 & 	$^2\!P^\circ$	 & 	$\frac{ 5}{2}$	 & 	$\frac{ 1}{2}$	 & 	$1.142(-1)^*$      	 & 	    41210.   	 & 	    62417.  \\
   4718.8\phantom{00000}	 & 	\llap{[\,}Ne\,{\sc iv}\,]	 & 	M1	 & 	$^2\!D^\circ$	 & 	$^2\!P^\circ$	 & 	$\frac{ 3}{2}$	 & 	$\frac{ 3}{2}$	 & 	$5.881(-1)^*$      	 & 	    41235.   	 & 	    62427.  \\
   4721.0\phantom{00000}	 & 	\llap{[\,}Ne\,{\sc iv}\,]	 & 	M1	 & 	$^2\!D^\circ$	 & 	$^2\!P^\circ$	 & 	$\frac{ 3}{2}$	 & 	$\frac{ 1}{2}$	 & 	$4.854(-1)^*$      	 & 	    41235.   	 & 	    62417.  \\
\hline
\noalign{\smallskip}
   1575.1\phantom{00000} & \llap{[\,}Ne\,{\sc v}\,]   & M1 &       $^{3}{\rm P}$ & $^{1}{\rm S}$       &           $1$ & $0$           &          &  $411.23$ & $63900.00$ \\
   1592.6\phantom{00000} & \llap{[\,}Ne\,{\sc v}\,]   & E2 &       $^{3}{\rm P}$ & $^{1}{\rm S}$       &           $2$ & $0$           &          & $1109.47$ & $63900.00$ \\
   2976.\phantom{000000}	 & 	\llap{[\,}Ne\,{\sc v}\,]	 & 	E2	 & 	$^1\!D$   	 & 	$^1\!S$   	 & 	$2$ 	 & 	$0$ 	 & 	$2.834$      	 & 	    30294.00 	 & 	    63900.00\\
   3301.0\phantom{00000}	 & 	\llap{[\,}Ne\,{\sc v}\,]	 & 	E2	 & 	$^3\!P$   	 & 	$^1\!D$   	 & 	$0$ 	 & 	$2$ 	 & 	$5.460(-5)^*$      	 & 	        0.00 	 & 	    30294.00\\
   3346.4\phantom{00000}	 & 	\llap{[\,}Ne\,{\sc v}\,]	 & 	M1	 & 	$^3\!P$   	 & 	$^1\!D$   	 & 	$1$ 	 & 	$2$ 	 & 	$1.222(-1)^*$      	 & 	      411.23 	 & 	    30294.00\\
   3426.5\phantom{00000}	 & 	\llap{[\,}Ne\,{\sc v}\,]	 & 	M1	 & 	$^3\!P$   	 & 	$^1\!D$   	 & 	$2$ 	 & 	$2$ 	 & 	$3.504(-1)^*$      	 & 	     1109.47 	 & 	    30294.00\\
  90133.3\phantom{00000}	 & 	\llap{[\,}Ne\,{\sc v}\,]	 & 	E2	 & 	$^3\!P$   	 & 	$^3\!P$   	 & 	$0$ 	 & 	$2$ 	 & 	$$      	 & 	        0.00 	 & 	     1109.47\\
 143216.8\phantom{00000}	 & 	\llap{[\,}Ne\,{\sc v}\,]	 & 	M1	 & 	$^3\!P$   	 & 	$^3\!P$   	 & 	$1$ 	 & 	$2$ 	 & 	$4.585(-3)^*$	 & 	      411.23 	 & 	     1109.47\\
 243175.\phantom{000000}	 & 	\llap{[\,}Ne\,{\sc v}\,]	 & 	M1	 & 	$^3\!P$   	 & 	$^3\!P$   	 & 	$0$ 	 & 	$1$ 	 & 	$1.666(-3)^*$	 & 	        0.00 	 & 	      411.23\\
\hline
\end{tabular}
}
\normalsize
\rm
\caption{ Lines of Ne ions.}\label{t1}
\end{center}
\end{table}

\newpage

% ------------------------------------------------------------------------------
%
% ------------------------------------------------------------------------------

\begin{table}[h!!!!!]
\begin{center}
\scriptsize{
\begin{tabular}{rlcr@{ -- }lr@{ -- }llr@{ -- }l}
\hline
\noalign{\smallskip}
$\lambda_{\rm vac}$ (\AA) \phantom{00}&Ion&TT&\multicolumn{2}{c}{Terms}&$J$&$J$&\multicolumn{1}{c}{$A_{\rm ki}$ (s$^{-1}$)}&\multicolumn{2}{c}{Levels (cm$^{-1}$)}\\
\hline
\noalign{\smallskip}
   4590.5464\phantom{00}	 & 	\llap{[\,}S\,{\sc i}\,]	 & 	M1	 & 	$^3\!P$   	 & 	$^1\!S$   	 & 	$1$ 	 & 	$0$ 	 & 	$$      	 & 	      396.06 	 & 	    22179.95\\
   7727.172\phantom{000}	 & 	\llap{[\,}S\,{\sc i}\,]	 & 	E2	 & 	$^1\!D$   	 & 	$^1\!S$   	 & 	$2$ 	 & 	$0$ 	 & 	$$      	 & 	     9238.61 	 & 	    22179.95\\
  10824.140\phantom{000}	 & 	\llap{[\,}S\,{\sc i}\,]	 & 	M1	 & 	$^3\!P$   	 & 	$^1\!D$   	 & 	$2$ 	 & 	$2$ 	 & 	$$      	 & 	        0.00 	 & 	     9238.61\\
  11308.950\phantom{000}	 & 	\llap{[\,}S\,{\sc i}\,]	 & 	M1	 & 	$^3\!P$   	 & 	$^1\!D$   	 & 	$1$ 	 & 	$2$ 	 & 	$$      	 & 	      396.06 	 & 	     9238.61\\
  11540.722\phantom{000}	 & 	\llap{[\,}S\,{\sc i}\,]	 & 	E2	 & 	$^3\!P$   	 & 	$^1\!D$   	 & 	$0$ 	 & 	$2$ 	 & 	$$      	 & 	      573.64 	 & 	     9238.61\\
 174325.4\phantom{00000}	 & 	\llap{[\,}S\,{\sc i}\,]	 & 	E2	 & 	$^3\!P$   	 & 	$^3\!P$   	 & 	$2$ 	 & 	$0$ 	 & 	$$      	 & 	        0.00 	 & 	      573.64\\
 252490.\phantom{000000}	 & 	\llap{[\,}S\,{\sc i}\,]	 & 	M1	 & 	$^3\!P$   	 & 	$^3\!P$   	 & 	$2$ 	 & 	$1$ 	 & 	$1.40(-3)$	 & 	        0.00 	 & 	      396.06\\
 563111.\phantom{000000}	 & 	\llap{[\,}S\,{\sc i}\,]	 & 	M1	 & 	$^3\!P$   	 & 	$^3\!P$   	 & 	$1$ 	 & 	$0$ 	 & 	$3.02(-4)$	 & 	      396.06 	 & 	      573.64\\
\hline
\noalign{\smallskip}
   1250.5845\phantom{00} & S \,{\sc ii}               & E1 & $^{4}{\rm S}^\circ$ & $^{4}{\rm P}$       & $\frac{3}{2}$ & $\frac{1}{2}$ & $4.43(+7)$ &   $0.00$ & $79962.61$ \\
   1253.8111\phantom{00} & S \,{\sc ii}               & E1 & $^{4}{\rm S}^\circ$ & $^{4}{\rm P}$       & $\frac{3}{2}$ & $\frac{3}{2}$ & $4.40(+7)$ &   $0.00$ & $79756.83$ \\
   1259.5190\phantom{00} & S \,{\sc ii}               & E1 & $^{4}{\rm S}^\circ$ & $^{4}{\rm P}$       & $\frac{3}{2}$ & $\frac{5}{2}$ & $4.34(+7)$ &   $0.00$ & $79395.39$ \\
   4069.749\phantom{000}	 & 	\llap{[\,}S\,{\sc ii}\,]	 & 	M1	 & 	$^4\!S^\circ$	 & 	$^2\!P^\circ$	 & 	$\frac{ 3}{2}$	 & 	$\frac{ 3}{2}$	 & 	$2.670(-1)^*$      	 & 	        0.00 	 & 	    24571.54\\
   4077.500\phantom{000}	 & 	\llap{[\,}S\,{\sc ii}\,]	 & 	M1	 & 	$^4\!S^\circ$	 & 	$^2\!P^\circ$	 & 	$\frac{ 3}{2}$	 & 	$\frac{ 1}{2}$	 & 	$1.076(-1)^*$      	 & 	        0.00 	 & 	    24524.83\\
   6718.29\phantom{0000}	 & 	\llap{[\,}S\,{\sc ii}\,]	 & 	E2	 & 	$^4\!S^\circ$	 & 	$^2\!D^\circ$	 & 	$\frac{ 3}{2}$	 & 	$\frac{ 5}{2}$	 & 	$3.338(-4)^*$      	 & 	        0.00 	 & 	    14884.73\\
   6732.67\phantom{0000}	 & 	\llap{[\,}S\,{\sc ii}\,]	 & 	E2	 & 	$^4\!S^\circ$	 & 	$^2\!D^\circ$	 & 	$\frac{ 3}{2}$	 & 	$\frac{ 3}{2}$	 & 	$1.231(-3)^*$      	 & 	        0.00 	 & 	    14852.94\\
  10289.55\phantom{0000}	 & 	\llap{[\,}S\,{\sc ii}\,]	 & 	M1	 & 	$^2\!D^\circ$	 & 	$^2\!P^\circ$	 & 	$\frac{ 3}{2}$	 & 	$\frac{ 3}{2}$	 & 	$1.644(-1)^*$      	 & 	    14852.94 	 & 	    24571.54\\
  10323.32\phantom{0000}	 & 	\llap{[\,}S\,{\sc ii}\,]	 & 	M1	 & 	$^2\!D^\circ$	 & 	$^2\!P^\circ$	 & 	$\frac{ 5}{2}$	 & 	$\frac{ 3}{2}$	 & 	$1.938(-1)^*$      	 & 	    14884.73 	 & 	    24571.54\\
  10339.24\phantom{0000}	 & 	\llap{[\,}S\,{\sc ii}\,]	 & 	M1	 & 	$^2\!D^\circ$	 & 	$^2\!P^\circ$	 & 	$\frac{ 3}{2}$	 & 	$\frac{ 1}{2}$	 & 	$1.812(-1)^*$      	 & 	    14852.94 	 & 	    24524.83\\
  10373.34\phantom{0000}	 & 	\llap{[\,}S\,{\sc ii}\,]	 & 	E2	 & 	$^2\!D^\circ$	 & 	$^2\!P^\circ$	 & 	$\frac{ 5}{2}$	 & 	$\frac{ 1}{2}$	 & 	$7.506(-2)^*$      	 & 	    14884.73 	 & 	    24524.83\\
2141000.\phantom{000000}	 & 	\llap{[\,}S\,{\sc ii}\,]	 & 	M1	 & 	$^2\!P^\circ$	 & 	$^2\!P^\circ$	 & 	$\frac{ 1}{2}$	 & 	$\frac{ 3}{2}$	 & 	$9.16(-7)$	 & 	    24524.83 	 & 	    24571.54\\
3146000.\phantom{000000}	 & 	\llap{[\,}S\,{\sc ii}\,]	 & 	M1	 & 	$^2\!D^\circ$	 & 	$^2\!D^\circ$	 & 	$\frac{ 3}{2}$	 & 	$\frac{ 5}{2}$	 & 	$3.452(-7)^*$	 & 	    14852.94 	 & 	    14884.73\\
\hline
\noalign{\smallskip}
   1704.3928\phantom{00} & \llap{[\,}S \,{\sc iii}\,] & M2 &       $^{3}{\rm P}$ & $^{5}{\rm S}^\circ$ &           $0$ & $2$           &          &   $0.00$ & $58671.92$ \\
   1713.1137\phantom{00} & S \,{\sc iii}\,]           & E1 &       $^{3}{\rm P}$ & $^{5}{\rm S}^\circ$ &           $1$ & $2$           &          & $298.68$ & $58671.92$ \\
   1728.9415\phantom{00} & S \,{\sc iii}\,]           & E1 &       $^{3}{\rm P}$ & $^{5}{\rm S}^\circ$ &           $2$ & $2$           &          & $833.06$ & $58671.92$ \\
   3722.69\phantom{0000}	 & 	\llap{[\,}S\,{\sc iii}\,]	 & 	M1	 & 	$^3\!P$   	 & 	$^1\!S$   	 & 	$1$ 	 & 	$0$ 	 & 	$1.016$      	 & 	      298.68 	 & 	    27161.00\\
   3798.25\phantom{0000} & \llap{[\,}S \,{\sc iii}\,] & E2 &       $^{3}{\rm P}$ & $^{1}{\rm S}$       &           $2$ & $0$           &          & $833.06$ & $27161.00$ \\
   6313.8\phantom{00000}	 & 	\llap{[\,}S\,{\sc iii}\,]	 & 	E2	 & 	$^1\!D$   	 & 	$^1\!S$   	 & 	$2$ 	 & 	$0$ 	 & 	$3.045$      	 & 	    11322.70 	 & 	    27161.00\\
   8831.8\phantom{00000}	 & 	\llap{[\,}S\,{\sc iii}\,]	 & 	E2	 & 	$^3\!P$   	 & 	$^1\!D$   	 & 	$0$ 	 & 	$2$ 	 & 	$2.570(-4)^*$      	 & 	        0.00 	 & 	    11322.70\\
   9071.1\phantom{00000}	 & 	\llap{[\,}S\,{\sc iii}\,]	 & 	M1	 & 	$^3\!P$   	 & 	$^1\!D$   	 & 	$1$ 	 & 	$2$ 	 & 	$3.107(-2)^*$      	 & 	      298.68 	 & 	    11322.70\\
   9533.2\phantom{00000}	 & 	\llap{[\,}S\,{\sc iii}\,]	 & 	M1	 & 	$^3\!P$   	 & 	$^1\!D$   	 & 	$2$ 	 & 	$2$ 	 & 	$1.856(-1)^*$      	 & 	      833.06 	 & 	    11322.70\\
 120038.8\phantom{00000}	 & 	\llap{[\,}S\,{\sc iii}\,]	 & 	E2	 & 	$^3\!P$   	 & 	$^3\!P$   	 & 	$0$ 	 & 	$2$ 	 & 	$5.008(-8)^*$      	 & 	        0.00 	 & 	      833.06\\
 187130.3\phantom{00000}	 & 	\llap{[\,}S\,{\sc iii}\,]	 & 	M1	 & 	$^3\!P$   	 & 	$^3\!P$   	 & 	$1$ 	 & 	$2$ 	 & 	$1.877(-3)^*$	 & 	      298.68 	 & 	      833.06\\
 334810.\phantom{000000}	 & 	\llap{[\,}S\,{\sc iii}\,]	 & 	M1	 & 	$^3\!P$   	 & 	$^3\!P$   	 & 	$0$ 	 & 	$1$ 	 & 	$8.429(-4)^*$	 & 	        0.00 	 & 	      298.68\\
\hline
\noalign{\smallskip}
   1387.459\phantom{000} & \llap{[\,}S \,{\sc iv}\,]  & M2 & $^{2}{\rm P}^\circ$ & $^{4}{\rm P}$       & $\frac{1}{2}$ & $\frac{5}{2}$ &          &   $0.00$ & $72074.20$ \\
   1398.050\phantom{000} & S \,{\sc iv}\,]            & E1 & $^{2}{\rm P}^\circ$ & $^{4}{\rm P}$       & $\frac{1}{2}$ & $\frac{3}{2}$ &          &   $0.00$ & $71528.20$ \\
   1404.800\phantom{000} & S \,{\sc iv}\,]            & E1 & $^{2}{\rm P}^\circ$ & $^{4}{\rm P}$       & $\frac{1}{2}$ & $\frac{1}{2}$ &          &   $0.00$ & $71184.50$ \\
   1406.019\phantom{000} & S \,{\sc iv}\,]            & E1 & $^{2}{\rm P}^\circ$ & $^{4}{\rm P}$       & $\frac{3}{2}$ & $\frac{5}{2}$ &          & $951.43$ & $72074.20$ \\
   1416.897\phantom{000} & S \,{\sc iv}\,]            & E1 & $^{2}{\rm P}^\circ$ & $^{4}{\rm P}$       & $\frac{3}{2}$ & $\frac{3}{2}$ &          & $951.43$ & $71528.20$ \\
   1423.831\phantom{000} & S \,{\sc iv}\,]            & E1 & $^{2}{\rm P}^\circ$ & $^{4}{\rm P}$       & $\frac{3}{2}$ & $\frac{1}{2}$ &          & $951.43$ & $71184.50$ \\
 105104.9\phantom{00000}	 & 	\llap{[\,}S\,{\sc iv}\,]	 & 	M1	 & 	$^2\!P^\circ$	 & 	$^2\!P^\circ$	 & 	$\frac{ 1}{2}$	 & 	$\frac{ 3}{2}$	 & 	$7.750(-3)^*$	 & 	        0.00 	 & 	      951.43\\
 112400.\phantom{000000} & \llap{[\,}S \,{\sc iv}\,]  & E2 &       $^{4}{\rm P}$ & $^{4}{\rm P}$       & $\frac{1}{2}$ & $\frac{5}{2}$ &          & $71184.50$ & $72074.20$ \\
 183200.\phantom{000000} & \llap{[\,}S \,{\sc iv}\,]  & M1 &       $^{4}{\rm P}$ & $^{4}{\rm P}$       & $\frac{3}{2}$ & $\frac{5}{2}$ & $2.63(-3)$ & $71528.20$ & $72074.20$ \\
 291000.\phantom{000000} & \llap{[\,}S \,{\sc iv}\,]  & M1 &       $^{4}{\rm P}$ & $^{4}{\rm P}$       & $\frac{1}{2}$ & $\frac{3}{2}$ & $9.13(-4)$ & $71184.50$ & $71528.20$ \\
\hline
\noalign{\smallskip}
   1188.281\phantom{000} & \llap{[\,}S \,{\sc v}\,] & M2 &       $^{1}{\rm S}$ & $^{3}{\rm P}^\circ$ & $0$ & $2$ &          &     $0.00$ & $84155.20$ \\
  88400.\phantom{000000} & \llap{[\,}S \,{\sc v}\,] & E2 & $^{3}{\rm P}^\circ$ & $^{3}{\rm P}^\circ$ & $0$ & $2$ &          & $83024.00$ & $84155.20$ \\
 131290.\phantom{000000} & \llap{[\,}S \,{\sc v}\,] & M1 & $^{3}{\rm P}^\circ$ & $^{3}{\rm P}^\circ$ & $1$ & $2$ & $5.96(-3)$ & $83393.50$ & $84155.20$ \\
 270600.\phantom{000000} & \llap{[\,}S \,{\sc v}\,] & M1 & $^{3}{\rm P}^\circ$ & $^{3}{\rm P}^\circ$ & $0$ & $1$ & $9.07(-4)$ & $83024.00$ & $83393.50$ \\
\hline
\end{tabular}
}
\normalsize
\rm
\caption{ Lines of S ions.}\label{t1}
\end{center}
\end{table}

%\newpage

% ------------------------------------------------------------------------------
%
% ------------------------------------------------------------------------------

\begin{table}[h!]
\begin{center}
\scriptsize{
\begin{tabular}{rlcr@{ -- }lr@{ -- }llr@{ -- }l}
\hline
\noalign{\smallskip}
$\lambda_{\rm vac}$ (\AA) \phantom{00}&Ion&TT&\multicolumn{2}{c}{Terms}&$J$&$J$&\multicolumn{1}{c}{$A_{\rm ki}$ (s$^{-1}$)}&\multicolumn{2}{c}{Levels (cm$^{-1}$)}\\
\hline
\noalign{\smallskip}
 113333.52\phantom{0000}	 & 	\llap{[\,}Cl\,{\sc i}\,]	 & 	M1	 & 	$^2\!P^\circ$	 & 	$^2\!P^\circ$	 & 	$\frac{ 3}{2}$	 & 	$\frac{ 1}{2}$	 & 	$1.24(-2)$	 & 	        0.00 	 & 	      882.35\\
\hline
\noalign{\smallskip}
   3587.055\phantom{000} & \llap{[\,}Cl\,{\sc ii}\,]  & E2 &       $^{3}{\rm P}$ & $^{1}{\rm S}$       &           $2$ & $0$           &          &    $0.00$ & $27878.02$ \\
   3678.902\phantom{000}	 & 	\llap{[\,}Cl\,{\sc ii}\,]	 & 	M1	 & 	$^3\!P$   	 & 	$^1\!S$   	 & 	$1$ 	 & 	$0$ 	 & 	$1.297$      	 & 	      696.00 	 & 	    27878.02\\
   6163.54\phantom{0000}	 & 	\llap{[\,}Cl\,{\sc ii}\,]	 & 	E2	 & 	$^1\!D$   	 & 	$^1\!S$   	 & 	$2$ 	 & 	$0$ 	 & 	$2.416$      	 & 	    11653.58 	 & 	    27878.02\\
   8581.05\phantom{0000}	 & 	\llap{[\,}Cl\,{\sc ii}\,]	 & 	M1	 & 	$^3\!P$   	 & 	$^1\!D$   	 & 	$2$ 	 & 	$2$ 	 & 	$1.038(-1)^*$      	 & 	        0.00 	 & 	    11653.58\\
   9126.10\phantom{0000}	 & 	\llap{[\,}Cl\,{\sc ii}\,]	 & 	M1	 & 	$^3\!P$   	 & 	$^1\!D$   	 & 	$1$ 	 & 	$2$ 	 & 	$2.872(-2)^*$      	 & 	      696.00 	 & 	    11653.58\\
   9383.41\phantom{0000}	 & 	\llap{[\,}Cl\,{\sc ii}\,]	 & 	E2	 & 	$^3\!P$   	 & 	$^1\!D$   	 & 	$0$ 	 & 	$2$ 	 & 	$1.134(-5)^*$      	 & 	      996.47 	 & 	    11653.58\\
 100354.\phantom{000000}	 & 	\llap{[\,}Cl\,{\sc ii}\,]	 & 	E2	 & 	$^3\!P$   	 & 	$^3\!P$   	 & 	$2$ 	 & 	$0$ 	 & 	$6.053(-7)^*$      	 & 	        0.00 	 & 	      996.47\\
 143678.\phantom{000000}	 & 	\llap{[\,}Cl\,{\sc ii}\,]	 & 	M1	 & 	$^3\!P$   	 & 	$^3\!P$   	 & 	$2$ 	 & 	$1$ 	 & 	$7.566(-3)^*$	 & 	        0.00 	 & 	      696.00\\
 332810.\phantom{000000}	 & 	\llap{[\,}Cl\,{\sc ii}\,]	 & 	M1	 & 	$^3\!P$   	 & 	$^3\!P$   	 & 	$1$ 	 & 	$0$ 	 & 	$1.462(-3)^*$	 & 	      696.00 	 & 	      996.47\\
\hline
\noalign{\smallskip}
   3343.81\phantom{0000}	 & 	\llap{[\,}Cl\,{\sc iii}\,]	 & 	M1	 & 	$^4\!S^\circ$	 & 	$^2\!P^\circ$	 & 	$\frac{ 3}{2}$	 & 	$\frac{ 3}{2}$	 & 	$$      	 & 	        0.00 	 & 	    29906.00\\
   3354.17\phantom{0000}	 & 	\llap{[\,}Cl\,{\sc iii}\,]	 & 	M1	 & 	$^4\!S^\circ$	 & 	$^2\!P^\circ$	 & 	$\frac{ 3}{2}$	 & 	$\frac{ 1}{2}$	 & 	$$      	 & 	        0.00 	 & 	    29813.60\\
   5519.25\phantom{0000}	 & 	\llap{[\,}Cl\,{\sc iii}\,]	 & 	E2	 & 	$^4\!S^\circ$	 & 	$^2\!D^\circ$	 & 	$\frac{ 3}{2}$	 & 	$\frac{ 5}{2}$	 & 	$$      	 & 	        0.00 	 & 	    18118.40\\
   5539.43\phantom{0000}	 & 	\llap{[\,}Cl\,{\sc iii}\,]	 & 	E2	 & 	$^4\!S^\circ$	 & 	$^2\!D^\circ$	 & 	$\frac{ 3}{2}$	 & 	$\frac{ 3}{2}$	 & 	$$      	 & 	        0.00 	 & 	    18052.40\\
   8436.3\phantom{00000}	 & 	\llap{[\,}Cl\,{\sc iii}\,]	 & 	M1	 & 	$^2\!D^\circ$	 & 	$^2\!P^\circ$	 & 	$\frac{ 3}{2}$	 & 	$\frac{ 3}{2}$	 & 	$$      	 & 	    18052.40 	 & 	    29906.00\\
   8483.5\phantom{00000}	 & 	\llap{[\,}Cl\,{\sc iii}\,]	 & 	M1	 & 	$^2\!D^\circ$	 & 	$^2\!P^\circ$	 & 	$\frac{ 5}{2}$	 & 	$\frac{ 3}{2}$	 & 	$$      	 & 	    18118.40 	 & 	    29906.00\\
   8502.5\phantom{00000}	 & 	\llap{[\,}Cl\,{\sc iii}\,]	 & 	M1	 & 	$^2\!D^\circ$	 & 	$^2\!P^\circ$	 & 	$\frac{ 3}{2}$	 & 	$\frac{ 1}{2}$	 & 	$$      	 & 	    18052.40 	 & 	    29813.60\\
   8550.5\phantom{00000}	 & 	\llap{[\,}Cl\,{\sc iii}\,]	 & 	E2	 & 	$^2\!D^\circ$	 & 	$^2\!P^\circ$	 & 	$\frac{ 5}{2}$	 & 	$\frac{ 1}{2}$	 & 	$$      	 & 	    18118.40 	 & 	    29813.60\\
1082000.\phantom{000000}	 & 	\llap{[\,}Cl\,{\sc iii}\,]	 & 	M1	 & 	$^2\!P^\circ$	 & 	$^2\!P^\circ$	 & 	$\frac{ 1}{2}$	 & 	$\frac{ 3}{2}$	 & 	$7.09(-6)$	 & 	    29813.60 	 & 	    29906.00\\
1515000.\phantom{000000}	 & 	\llap{[\,}Cl\,{\sc iii}\,]	 & 	M1	 & 	$^2\!D^\circ$	 & 	$^2\!D^\circ$	 & 	$\frac{ 3}{2}$	 & 	$\frac{ 5}{2}$	 & 	$3.10(-6)$	 & 	    18052.40 	 & 	    18118.40\\
\hline
\noalign{\smallskip}
   3119.51\phantom{0000}	 & 	\llap{[\,}Cl\,{\sc iv}\,]	 & 	M1	 & 	$^3\!P$   	 & 	$^1\!S$   	 & 	$1$ 	 & 	$0$ 	 & 	$$      	 & 	      492.50 	 & 	    32548.80\\
   3204.52\phantom{0000} & \llap{[\,}Cl\,{\sc iv}\,]  & E2 &       $^{3}{\rm P}$ & $^{1}{\rm S}$       &           $2$ & $0$           &          & $1342.90$ & $32548.80$ \\
   5324.47\phantom{0000}	 & 	\llap{[\,}Cl\,{\sc iv}\,]	 & 	E2	 & 	$^1\!D$   	 & 	$^1\!S$   	 & 	$2$ 	 & 	$0$ 	 & 	$$      	 & 	    13767.60 	 & 	    32548.80\\
   7263.4\phantom{00000}	 & 	\llap{[\,}Cl\,{\sc iv}\,]	 & 	E2	 & 	$^3\!P$   	 & 	$^1\!D$   	 & 	$0$ 	 & 	$2$ 	 & 	$$      	 & 	        0.00 	 & 	    13767.60\\
   7532.9\phantom{00000}	 & 	\llap{[\,}Cl\,{\sc iv}\,]	 & 	M1	 & 	$^3\!P$   	 & 	$^1\!D$   	 & 	$1$ 	 & 	$2$ 	 & 	$$      	 & 	      492.50 	 & 	    13767.60\\
   8048.5\phantom{00000}	 & 	\llap{[\,}Cl\,{\sc iv}\,]	 & 	M1	 & 	$^3\!P$   	 & 	$^1\!D$   	 & 	$2$ 	 & 	$2$ 	 & 	$$      	 & 	     1342.90 	 & 	    13767.60\\
  74470.\phantom{000000}	 & 	\llap{[\,}Cl\,{\sc iv}\,]	 & 	E2	 & 	$^3\!P$   	 & 	$^3\!P$   	 & 	$0$ 	 & 	$2$ 	 & 	$$      	 & 	        0.00 	 & 	     1342.90\\
 117590.\phantom{000000}	 & 	\llap{[\,}Cl\,{\sc iv}\,]	 & 	M1	 & 	$^3\!P$   	 & 	$^3\!P$   	 & 	$1$ 	 & 	$2$ 	 & 	$8.29(-3)$	 & 	      492.50 	 & 	     1342.90\\
 203000.\phantom{000000}	 & 	\llap{[\,}Cl\,{\sc iv}\,]	 & 	M1	 & 	$^3\!P$   	 & 	$^3\!P$   	 & 	$0$ 	 & 	$1$ 	 & 	$2.15(-3)$	 & 	        0.00 	 & 	      492.50\\
\hline
\noalign{\smallskip}
  67090.\phantom{000000}	 & 	\llap{[\,}Cl\,{\sc v}\,]	 & 	M1	 & 	$^2\!P^\circ$	 & 	$^2\!P^\circ$	 & 	$\frac{ 1}{2}$	 & 	$\frac{ 3}{2}$	 & 	$2.98(-2)$	 & 	        0.00 	 & 	     1490.50\\
  72400.\phantom{000000}	 & 	\llap{[\,}Cl\,{\sc v}\,]	 & 	E2	 & 	$^4\!P$   	 & 	$^4\!P$   	 & 	$\frac{ 1}{2}$	 & 	$\frac{ 5}{2}$	 & 	$$      	 & 	    86000.+v 	 & 	    87381.+v\\
 118600.\phantom{000000}	 & 	\llap{[\,}Cl\,{\sc v}\,]	 & 	M1	 & 	$^4\!P$   	 & 	$^4\!P$   	 & 	$\frac{ 3}{2}$	 & 	$\frac{ 5}{2}$	 & 	$9.70(-3)$	 & 	    86538.+v 	 & 	    87381.+v\\
 186000.\phantom{000000}	 & 	\llap{[\,}Cl\,{\sc v}\,]	 & 	M1	 & 	$^4\!P$   	 & 	$^4\!P$   	 & 	$\frac{ 1}{2}$	 & 	$\frac{ 3}{2}$	 & 	$3.50(-3)$	 & 	    86000.+v 	 & 	    86538.+v\\
\hline
\noalign{\smallskip}
  58200.\phantom{000000} & \llap{[\,}Cl\,{\sc vi}\,] & E2 & $^{3}{\rm P}^\circ$ & $^{3}{\rm P}^\circ$ & $0$ & $2$ &          & $97405.+x$ & $99123.+x$ \\
  85800.\phantom{000000} & \llap{[\,}Cl\,{\sc vi}\,] & M1 & $^{3}{\rm P}^\circ$ & $^{3}{\rm P}^\circ$ & $1$ & $2$ & $2.13(-2)$ & $97958.+x$ & $99123.+x$ \\
 180800.\phantom{000000} & \llap{[\,}Cl\,{\sc vi}\,] & M1 & $^{3}{\rm P}^\circ$ & $^{3}{\rm P}^\circ$ & $0$ & $1$ & $3.04(-3)$ & $97405.+x$ & $97958.+x$ \\
\hline
\end{tabular}
}
\normalsize
\rm
\caption{ Lines of Cl ions.}\label{t1}
\end{center}
\end{table}
																		
%\newpage

% ------------------------------------------------------------------------------
%
% ------------------------------------------------------------------------------

\clearpage
\begin{table}[h!]
\begin{center}
\scriptsize{
\begin{tabular}{rlcr@{ -- }lr@{ -- }llr@{ -- }l}
\hline
\noalign{\smallskip}
$\lambda_{\rm vac}$ (\AA) \phantom{00}&Ion&TT&\multicolumn{2}{c}{Terms}&$J$&$J$&\multicolumn{1}{c}{$A_{\rm ki}$ (s$^{-1}$)}&\multicolumn{2}{c}{Levels (cm$^{-1}$)}\\
\hline
\noalign{\smallskip}
  69852.74\phantom{0000}	 & 	\llap{[\,}Ar\,{\sc ii}\,]	 & 	M1	 & 	$^2\!P^\circ$	 & 	$^2\!P^\circ$	 & 	$\frac{ 3}{2}$	 & 	$\frac{ 1}{2}$	 & 	$5.28(-2)$	 & 	        0.00 	 & 	     1431.58\\
\hline
\noalign{\smallskip}
   3006.10\phantom{0000} & \llap{[\,}Ar\,{\sc iii}\,] & E2 &       $^{3}{\rm P}$ & $^{1}{\rm S}$       &           $2$ & $0$           &          &    $0.00$ & $33265.70$ \\
   3110.08\phantom{0000}	 & 	\llap{[\,}Ar\,{\sc iii}\,]	 & 	M1	 & 	$^3\!P$   	 & 	$^1\!S$   	 & 	$1$ 	 & 	$0$ 	 & 	$$      	 & 	     1112.18 	 & 	    33265.70\\
   5193.27\phantom{0000}	 & 	\llap{[\,}Ar\,{\sc iii}\,]	 & 	E2	 & 	$^1\!D$   	 & 	$^1\!S$   	 & 	$2$ 	 & 	$0$ 	 & 	$$      	 & 	    14010.00 	 & 	    33265.70\\
   7137.8\phantom{00000}	 & 	\llap{[\,}Ar\,{\sc iii}\,]	 & 	M1	 & 	$^3\!P$   	 & 	$^1\!D$   	 & 	$2$ 	 & 	$2$ 	 & 	$$      	 & 	        0.00 	 & 	    14010.00\\
   7753.2\phantom{00000}	 & 	\llap{[\,}Ar\,{\sc iii}\,]	 & 	M1	 & 	$^3\!P$   	 & 	$^1\!D$   	 & 	$1$ 	 & 	$2$ 	 & 	$$      	 & 	     1112.18 	 & 	    14010.00\\
   8038.8\phantom{00000}	 & 	\llap{[\,}Ar\,{\sc iii}\,]	 & 	E2	 & 	$^3\!P$   	 & 	$^1\!D$   	 & 	$0$ 	 & 	$2$ 	 & 	$$      	 & 	     1570.28 	 & 	    14010.00\\
  63682.9\phantom{00000}	 & 	\llap{[\,}Ar\,{\sc iii}\,]	 & 	E2	 & 	$^3\!P$   	 & 	$^3\!P$   	 & 	$2$ 	 & 	$0$ 	 & 	$$      	 & 	        0.00 	 & 	     1570.28\\
  89913.8\phantom{00000}	 & 	\llap{[\,}Ar\,{\sc iii}\,]	 & 	M1	 & 	$^3\!P$   	 & 	$^3\!P$   	 & 	$2$ 	 & 	$1$ 	 & 	$3.09(-2)$	 & 	        0.00 	 & 	     1112.18\\
 218291.\phantom{000000}	 & 	\llap{[\,}Ar\,{\sc iii}\,]	 & 	M1	 & 	$^3\!P$   	 & 	$^3\!P$   	 & 	$1$ 	 & 	$0$ 	 & 	$5.19(-3)$	 & 	     1112.18 	 & 	     1570.28\\
\hline
\noalign{\smallskip}
   2854.48\phantom{0000}	 & 	\llap{[\,}Ar\,{\sc iv}\,]	 & 	M1	 & 	$^4\!S^\circ$	 & 	$^2\!P^\circ$	 & 	$\frac{ 3}{2}$	 & 	$\frac{ 3}{2}$	 & 	$2.134$      	 & 	        0.00 	 & 	    35032.60\\
   2868.99\phantom{0000}	 & 	\llap{[\,}Ar\,{\sc iv}\,]	 & 	M1	 & 	$^4\!S^\circ$	 & 	$^2\!P^\circ$	 & 	$\frac{ 3}{2}$	 & 	$\frac{ 1}{2}$	 & 	$8.702(-1)^*$      	 & 	        0.00 	 & 	    34855.50\\
   4712.69\phantom{0000}	 & 	\llap{[\,}Ar\,{\sc iv}\,]	 & 	E2	 & 	$^4\!S^\circ$	 & 	$^2\!D^\circ$	 & 	$\frac{ 3}{2}$	 & 	$\frac{ 5}{2}$	 & 	$1.906(-3)^*$      	 & 	        0.00 	 & 	    21219.30\\
   4741.49\phantom{0000}	 & 	\llap{[\,}Ar\,{\sc iv}\,]	 & 	E2	 & 	$^4\!S^\circ$	 & 	$^2\!D^\circ$	 & 	$\frac{ 3}{2}$	 & 	$\frac{ 3}{2}$	 & 	$2.277(-2)^*$      	 & 	        0.00 	 & 	    21090.40\\
   7172.5\phantom{00000}	 & 	\llap{[\,}Ar\,{\sc iv}\,]	 & 	M1	 & 	$^2\!D^\circ$	 & 	$^2\!P^\circ$	 & 	$\frac{ 3}{2}$	 & 	$\frac{ 3}{2}$	 & 	$8.915(-1)^*$      	 & 	    21090.40 	 & 	    35032.60\\
   7239.4\phantom{00000}	 & 	\llap{[\,}Ar\,{\sc iv}\,]	 & 	M1	 & 	$^2\!D^\circ$	 & 	$^2\!P^\circ$	 & 	$\frac{ 5}{2}$	 & 	$\frac{ 3}{2}$	 & 	$6.559(-1)^*$      	 & 	    21219.30 	 & 	    35032.60\\
   7264.7\phantom{00000}	 & 	\llap{[\,}Ar\,{\sc iv}\,]	 & 	M1	 & 	$^2\!D^\circ$	 & 	$^2\!P^\circ$	 & 	$\frac{ 3}{2}$	 & 	$\frac{ 1}{2}$	 & 	$6.660(-1)^*$      	 & 	    21090.40 	 & 	    34855.50\\
   7333.4\phantom{00000}	 & 	\llap{[\,}Ar\,{\sc iv}\,]	 & 	E2	 & 	$^2\!D^\circ$	 & 	$^2\!P^\circ$	 & 	$\frac{ 5}{2}$	 & 	$\frac{ 1}{2}$	 & 	$1.179(-1)^*$      	 & 	    21219.30 	 & 	    34855.50\\
 564700.\phantom{000000}	 & 	\llap{[\,}Ar\,{\sc iv}\,]	 & 	M1	 & 	$^2\!P^\circ$	 & 	$^2\!P^\circ$	 & 	$\frac{ 1}{2}$	 & 	$\frac{ 3}{2}$	 & 	$4.937(-5)^*$	 & 	    34855.50 	 & 	    35032.60\\
 776000.\phantom{000000}	 & 	\llap{[\,}Ar\,{\sc iv}\,]	 & 	M1	 & 	$^2\!D^\circ$	 & 	$^2\!D^\circ$	 & 	$\frac{ 3}{2}$	 & 	$\frac{ 5}{2}$	 & 	$2.287(-5)^*$	 & 	    21090.40 	 & 	    21219.30\\
\hline
\noalign{\smallskip}
   2691.63\phantom{0000}	 & 	\llap{[\,}Ar\,{\sc v}\,]	 & 	M1	 & 	$^3\!P$   	 & 	$^1\!S$   	 & 	$1$ 	 & 	$0$ 	 & 	$$      	 & 	      763.23 	 & 	    37915.50\\
   2786.55\phantom{0000} & \llap{[\,}Ar\,{\sc v}\,]   & E2 &       $^{3}{\rm P}$ & $^{1}{\rm S}$       &           $2$ & $0$           &          &   $2028.80$ &  $37915.50$ \\
   4626.22\phantom{0000}	 & 	\llap{[\,}Ar\,{\sc v}\,]	 & 	E2	 & 	$^1\!D$   	 & 	$^1\!S$   	 & 	$2$ 	 & 	$0$ 	 & 	$$      	 & 	    16299.60 	 & 	    37915.50\\
   6135.12\phantom{0000}	 & 	\llap{[\,}Ar\,{\sc v}\,]	 & 	E2	 & 	$^3\!P$   	 & 	$^1\!D$   	 & 	$0$ 	 & 	$2$ 	 & 	$$      	 & 	        0.00 	 & 	    16299.60\\
   6436.51\phantom{0000}	 & 	\llap{[\,}Ar\,{\sc v}\,]	 & 	M1	 & 	$^3\!P$   	 & 	$^1\!D$   	 & 	$1$ 	 & 	$2$ 	 & 	$$      	 & 	      763.23 	 & 	    16299.60\\
   7007.3\phantom{00000}	 & 	\llap{[\,}Ar\,{\sc v}\,]	 & 	M1	 & 	$^3\!P$   	 & 	$^1\!D$   	 & 	$2$ 	 & 	$2$ 	 & 	$$      	 & 	     2028.80 	 & 	    16299.60\\
  49290.2\phantom{00000}	 & 	\llap{[\,}Ar\,{\sc v}\,]	 & 	E2	 & 	$^3\!P$   	 & 	$^3\!P$   	 & 	$0$ 	 & 	$2$ 	 & 	$$      	 & 	        0.00 	 & 	     2028.80\\
  79015.8\phantom{00000}	 & 	\llap{[\,}Ar\,{\sc v}\,]	 & 	M1	 & 	$^3\!P$   	 & 	$^3\!P$   	 & 	$1$ 	 & 	$2$ 	 & 	$2.73(-2)$	 & 	      763.23 	 & 	     2028.80\\
 131021.9\phantom{00000}	 & 	\llap{[\,}Ar\,{\sc v}\,]	 & 	M1	 & 	$^3\!P$   	 & 	$^3\!P$   	 & 	$0$ 	 & 	$1$ 	 & 	$7.99(-3)$	 & 	        0.00 	 & 	      763.23\\
\hline
\noalign{\smallskip}
    978.58\phantom{0000} & \llap{[\,}Ar\,{\sc vi}\,]  & M2 & $^{2}{\rm P}^\circ$ & $^{4}{\rm P}$       & $\frac{1}{2}$ & $\frac{5}{2}$ &          &      $0.00$ & $102189.20$ \\
  45295.\phantom{000000} & \llap{[\,}Ar\,{\sc vi}\,] & M1 & $^{2}{\rm P}^\circ$ & $^{2}{\rm P}^\circ$ & $\frac{1}{2}$ & $\frac{3}{2}$ & $9.68(-2)$ & $0.00$ & $2207.74$ \\
  49237.\phantom{000000} & \llap{[\,}Ar\,{\sc vi}\,]  & E2 &       $^{4}{\rm P}$ & $^{4}{\rm P}$       & $\frac{1}{2}$ & $\frac{5}{2}$ &          & $100158.20$ & $102189.20$ \\
  81010.\phantom{000000} & \llap{[\,}Ar\,{\sc vi}\,]  & M1 &       $^{4}{\rm P}$ & $^{4}{\rm P}$       & $\frac{3}{2}$ & $\frac{5}{2}$ & $3.04(-2)$ & $100954.80$ & $102189.20$ \\
 125530.\phantom{000000} & \llap{[\,}Ar\,{\sc vi}\,]  & M1 &       $^{4}{\rm P}$ & $^{4}{\rm P}$       & $\frac{1}{2}$ & $\frac{3}{2}$ & $1.14(-2)$ & $100158.20$ & $100954.80$ \\
\hline
\noalign{\smallskip}
    865.16\phantom{0000} & \llap{[\,}Ar\,{\sc vii}\,] & M2 &       $^{1}{\rm S}$ & $^{3}{\rm P}^\circ$ &           $0$ & $2$           &          &      $0.00$ & $115585.00$ \\
  40310.\phantom{000000} & \llap{[\,}Ar\,{\sc vii}\,] & E2 & $^{3}{\rm P}^\circ$ & $^{3}{\rm P}^\circ$ &           $0$ & $2$           &          & $113104.00$ & $115585.00$ \\
  59500.\phantom{000000} & \llap{[\,}Ar\,{\sc vii}\,] & M1 & $^{3}{\rm P}^\circ$ & $^{3}{\rm P}^\circ$ &           $1$ & $2$           & $6.41(-2)$ & $113904.00$ & $115585.00$ \\
 125000.\phantom{000000} & \llap{[\,}Ar\,{\sc vii}\,] & M1 & $^{3}{\rm P}^\circ$ & $^{3}{\rm P}^\circ$ &           $0$ & $1$           & $9.21(-3)$ & $113104.00$ & $113904.00$ \\
\hline
\end{tabular}
}
\normalsize
\rm
\caption{ Lines of Ar ions.}\label{t1}
\end{center}
\end{table}

\

In Tables 1.2--1.8, it is easy to see which ions  are available for each element in a given wavelength range. For example, there is no line of N\,{\sc iii} in the optical. One can also see which lines arise from the same upper level, and therefore have intensity ratios that depend only on atomic physics and not on the conditions inside the nebula. For example, for [O\,{\sc iii}], this is the case for the  4969 and  5007\AA\ lines, whose intensity ratio is given by the ratio of transition probabilities divided by the ratio of wavelengths.

Tables 1.9--1.14 group the lines according to the electronic configuration of their ions. Fig. 1.6 shows simplified energy level diagrams, which allow one to understand which line pairs are temperature  or density indicators. Of course, in practice, one uses only the lines with largest transition probabilities. For example,  [O\,{\sc iii}] $\lambda$4363/4931 could be a temperature indicator as  [O\,{\sc iii}] $\lambda$4663/5007 but the intensity of the  $\lambda$4931 line is less than one  thousand's of that of $\lambda$5007 (as inferred from Table 1.4, comparing the transition probabilities).

% -----------------------------------------------------------
% --------------------- GRAPH 3 -----------------------------
% -----------------------------------------------------------
\begin{figure}[h!]
\centering
\includegraphics[scale=0.6]{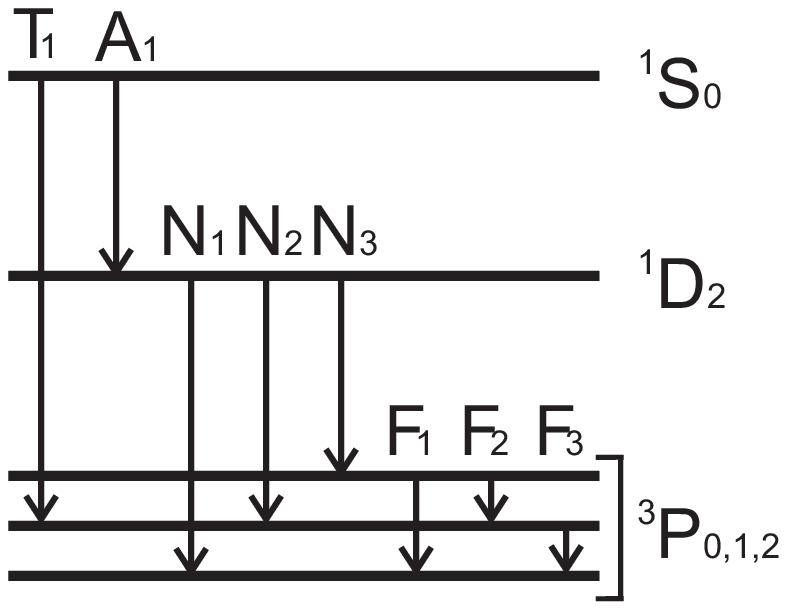}
\includegraphics[scale=0.6]{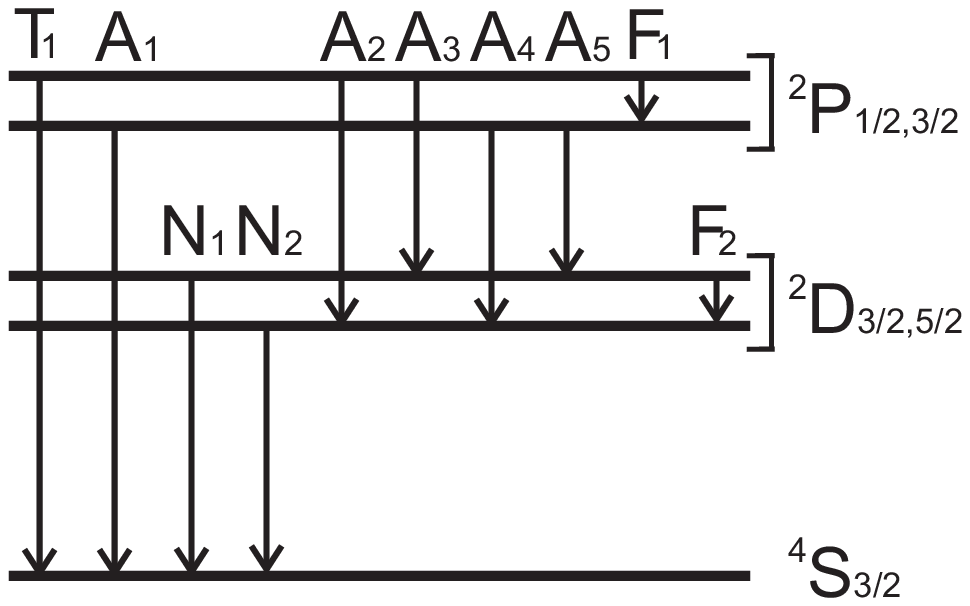}
\caption{\footnotesize Simplified energy level diagrams with transition types indicated. A stands for ``auroral'', N for ``nebular'', T for ``transauroral'' and F for ``fine structure"".  Left: ions p$^4$ (Table 1.10) and p$^2$ (Table 1.12). Right:  ions p$^3$ (Table 1.11) }
\label{Conf}
\end{figure}
% -----------------------------------------------------------
% -----------------------------------------------------------

\begin{table}[h!]
\begin{center}
\scriptsize{
\begin{tabular}{cccc}
\hline
\noalign{\smallskip}
 & 	\llap{[\,}Ne\,{\sc ii}\,]	  & 	\llap{[\,}Cl\,{\sc i}\,]	 & 	\llap{[\,}Ar\,{\sc ii}\,]	\\			
\noalign{\smallskip}
\hline
\noalign{\smallskip}
F$_1$ & 	 128135.48\phantom{0}	  & 	 113333.52\phantom{0}	 & 	  69852.74\phantom{0}	\\			
\hline
\end{tabular}
}
\normalsize
\rm
\caption{ Lines from 2p$^5$ and 3p$^5$ ions. The first column gives the type of line, as defined in Fig. 1.6. }\label{t1}
\end{center}
\end{table}

\begin{table}[h!]
\begin{center}
\scriptsize{
\begin{tabular}{cccccc}
\hline
\noalign{\smallskip}
 & 	\llap{[\,}O\,{\sc i}\,]	 & 	\llap{[\,}Ne\,{\sc iii}\,]	 & 	\llap{[\,}S\,{\sc i}\,]	 & 	\llap{[\,}Cl\,{\sc ii}\,]	 & 	\llap{[\,}Ar\,{\sc iii}\,]	\\			
\noalign{\smallskip}
\hline
\noalign{\smallskip}
T$_1$ & 	   2973.1538\phantom{0}	 & 	   1814.730\phantom{0}	 & 	   4590.5464\phantom{0}	 & 	   3678.902\phantom{0}	 & 	   3110.08\phantom{0}	\\			
A$_1$ & 	   5578.8874\phantom{0}	 & 	   3343.50\phantom{00}	 & 	   7727.172\phantom{00}	 & 	   6163.54\phantom{00}	 & 	   5193.27\phantom{0}	\\			
N$_1$ & 	   6302.046\phantom{00}	 & 	   3870.16\phantom{00}	 & 	  10824.140\phantom{00}	 & 	   8581.05\phantom{00}	 & 	   7137.8\phantom{00}	\\			
N$_2$ & 	   6365.536\phantom{00}	 & 	   3968.91\phantom{00}	 & 	  11308.950\phantom{00}	 & 	   9126.10\phantom{00}	 & 	   7753.2\phantom{00}	\\			
N$_3$ & 	   6393.500\phantom{00}	 & 	   4013.14\phantom{00}	 & 	  11540.722\phantom{00}	 & 	   9383.41\phantom{00}	 & 	   8038.8\phantom{00}	\\			
F$_1$ & 	 440573.\phantom{00000}	 & 	 108630.5\phantom{000}	 & 	 174325.4\phantom{0000}	 & 	 100354.\phantom{0000}	 & 	  63682.9\phantom{00}	\\			
F$_2$ & 	 631850.\phantom{00000}	 & 	 155550.5\phantom{000}	 & 	 252490.\phantom{00000}	 & 	 143678.\phantom{0000}	 & 	  89913.8\phantom{00}	\\			
F$_3$ & 	1455350.\phantom{00000}	 & 	 360135.\phantom{0000}	 & 	 563111.\phantom{00000}	 & 	 332810.\phantom{0000}	 & 	 218291.\phantom{000}	\\			
\hline
\end{tabular}
}
\normalsize
\rm
\caption{ Lines from 2p$^4$ and 3p$^4$ ions. The first column gives the type of line, as defined in Fig. 1.6.}\label{t1}
\end{center}
\end{table}

\begin{table}[h!]
\begin{center}
\scriptsize{
\begin{tabular}{ccccccc}
\hline
\noalign{\smallskip}
 & 	\llap{[\,}N\,{\sc i}\,]	 & 	\llap{[\,}O\,{\sc ii}\,]	 & 	\llap{[\,}Ne\,{\sc iv}\,]	 & 	\llap{[\,}S\,{\sc ii}\,]	 & 	\llap{[\,}Cl\,{\sc iii}\,]	 & 	\llap{[\,}Ar\,{\sc iv}\,]	\\			
\noalign{\smallskip}
\hline
\noalign{\smallskip}
T$_1$ & 	   3467.4898\phantom{0}	 & 	   2470.966\phantom{0}	 & 	   4713.2\phantom{0}	 & 	   4069.749\phantom{0}	 & 	   3343.81\phantom{0}	 & 	   2854.48\phantom{0}	\\			
T$_2$ & 	   3467.5362\phantom{0}	 & 	   2471.088\phantom{0}	 & 	   4715.4\phantom{0}	 & 	   4077.500\phantom{0}	 & 	   3354.17\phantom{0}	 & 	   2868.99\phantom{0}	\\			
N$_1$ & 	   5199.3490\phantom{0}	 & 	   3727.092\phantom{0}	 & 	   4718.8\phantom{0}	 & 	   6718.29\phantom{00}	 & 	   5519.25\phantom{0}	 & 	   4712.69\phantom{0}	\\			
N$_2$ & 	   5201.7055\phantom{0}	 & 	   3729.875\phantom{0}	 & 	   4721.0\phantom{0}	 & 	   6732.67\phantom{00}	 & 	   5539.43\phantom{0}	 & 	   4741.49\phantom{0}	\\			
A$_1$ & 	  10400.587\phantom{00}	 & 	   7320.94\phantom{00}	 & 		                     & 	  10289.55\phantom{00}	 & 	   8436.3\phantom{00}	 & 	   7172.5\phantom{00}	\\			
A$_2$ & 	  10401.004\phantom{00}	 & 	   7322.01\phantom{00}	 & 		                     & 	  10323.32\phantom{00}	 & 	   8483.5\phantom{00}	 & 	   7239.4\phantom{00}	\\			
A$_3$ & 	  10410.021\phantom{00}	 & 	   7331.68\phantom{00}	 & 		                     & 	  10339.24\phantom{00}	 & 	   8502.5\phantom{00}	 & 	   7264.7\phantom{00}	\\			
A$_4$ & 	  10410.439\phantom{00}	 & 	   7332.75\phantom{00}	 & 		                     & 	  10373.34\phantom{00}	 & 	   8550.5\phantom{00}	 & 	   7333.4\phantom{00}	\\			
F$_1$ & 		                     & 	4995000.\phantom{0000}	 & 		                     & 	2141000.\phantom{0000}	 & 	1082000.\phantom{000}	 & 	 564700.\phantom{000}	\\			
F$_2$ & 		                     & 		                     & 		                     & 	3146000.\phantom{0000}	 & 	1515000.\phantom{000}	 & 	 776000.\phantom{000}	\\			
\hline
\end{tabular}
}
\normalsize
\rm
\caption{ Lines from 2p$^3$ and 3p$^3$ ions. The first column gives the type of line, as defined in Fig. 1.6.}\label{t1}
\end{center}
\end{table}

\begin{table}[h!]
\begin{center}
\scriptsize{
\begin{tabular}{cccccccc}
\hline
\noalign{\smallskip}
 & 	\llap{[\,}C\,{\sc i}\,]	 & 	\llap{[\,}N\,{\sc ii}\,]	 & 	\llap{[\,}O\,{\sc iii}\,]	 & 	\llap{[\,}Ne\,{\sc v}\,]	 & 	\llap{[\,}S\,{\sc iii}\,]	 & 	\llap{[\,}Cl\,{\sc iv}\,]	 & 	\llap{[\,}Ar\,{\sc v}\,]	\\			
\noalign{\smallskip}
\hline
\noalign{\smallskip}
T$_1$ & 	   4622.864\phantom{0}	 & 	   3063.728\phantom{0}	 & 	   2321.664\phantom{0}	 & 	   1575.1\phantom{0}	 & 	   3722.69\phantom{0}	 & 	   3119.51\phantom{0}	 & 	   2691.63\phantom{0}	\\			
A$_1$ & 	   8729.52\phantom{00}	 & 	   5756.240\phantom{0}	 & 	   4364.436\phantom{0}	 & 	   2976.\phantom{00}	 & 	   6313.8\phantom{00}	 & 	   5324.47\phantom{0}	 & 	   4626.22\phantom{0}	\\			
N$_1$ & 	   9811.01\phantom{00}	 & 	   6529.04\phantom{00}	 & 	   4932.603\phantom{0}	 & 	   3301.0\phantom{0}	 & 	   8831.8\phantom{00}	 & 	   7263.4\phantom{00}	 & 	   6135.12\phantom{0}	\\			
N$_2$ & 	   9826.82\phantom{00}	 & 	   6549.85\phantom{00}	 & 	   4960.295\phantom{0}	 & 	   3346.4\phantom{0}	 & 	   9071.1\phantom{00}	 & 	   7532.9\phantom{00}	 & 	   6436.51\phantom{0}	\\			
N$_3$ & 	   9852.96\phantom{00}	 & 	   6585.28\phantom{00}	 & 	   5008.240\phantom{0}	 & 	   3426.5\phantom{0}	 & 	   9533.2\phantom{00}	 & 	   8048.5\phantom{00}	 & 	   7007.3\phantom{00}	\\			
F$_1$ & 	2304000.\phantom{0000}	 & 	 764500.\phantom{0000}	 & 	 326612.\phantom{0000}	 & 	  90133.3\phantom{0}	 & 	 120038.8\phantom{00}	 & 	  74470.\phantom{000}	 & 	  49290.2\phantom{00}	\\			
F$_2$ & 	3704000.\phantom{0000}	 & 	1217600.\phantom{0000}	 & 	 518145.\phantom{0000}	 & 	 143216.8\phantom{0}	 & 	 187130.3\phantom{00}	 & 	 117590.\phantom{000}	 & 	  79015.8\phantom{00}	\\			
F$_3$ & 	6100000.\phantom{0000}	 & 	2055000.\phantom{0000}	 & 	 883560.\phantom{0000}	 & 	 243175.\phantom{00}	 & 	 334810.\phantom{000}	 & 	 203000.\phantom{000}	 & 	 131021.9\phantom{00}	\\			
\hline
\end{tabular}
}
\normalsize
\rm
\caption{  Lines from 2p$^2$ and 3p$^2$ ions. The first column gives the type of line, as defined in Fig. 1.6.}\label{t1}
\end{center}
\end{table}

\begin{table}[h!]
\begin{center}
\scriptsize{
\begin{tabular}{cccccc}
\hline
\noalign{\smallskip}
 & 	\llap{[\,}C\,{\sc ii}\,]	 & 	\llap{[\,}N\,{\sc iii}\,]	 & 	\llap{[\,}O\,{\sc iv}\,]	 & 	\llap{[\,}S\,{\sc iv}\,]	 & 	\llap{[\,}Cl\,{\sc v}\,] 		\\			
\noalign{\smallskip}
\hline
\noalign{\smallskip}
F$_1$ & 	1576800.\phantom{0}	 & 	 573400.\phantom{0}	 & 	 258903.\phantom{0}	 & 	 105104.9\phantom{0}	 & 	  67090.\phantom{0}	 \\			
F$_2$ & 		                 & 		                 & 		                 & 		                     & 	  72400.\phantom{0}	 \\			
F$_3$ & 		                 & 		                 & 		                 & 		                     & 	 118600.\phantom{0}	 \\			
F$_4$ & 		                 & 		                 & 		                 & 		                     & 	 186000.\phantom{0}	 \\			
\hline
\end{tabular}
}
\normalsize
\rm
\caption{ Lines from 2p and 3p ions. The first column gives the type of line, as defined in Fig. 1.6.}\label{t1}
\end{center}
\end{table}

\begin{table}[h!]
\begin{center}
\scriptsize{
\begin{tabular}{cccc}
\hline
\noalign{\smallskip}
 & 	\llap{[\,}C\,{\sc iii}\,]	 & 	\llap{[\,}N\,{\sc iv}\,]	 & 	\llap{[\,}O\,{\sc v}\,]	 \\			
\noalign{\smallskip}
\hline
\noalign{\smallskip}
F$_1$ & 	1249200.\phantom{0}	 & 	 482900.\phantom{0}	 & 	 225800.\phantom{0}	\\			
F$_2$ & 	1774300.\phantom{0}	 & 	 694000.\phantom{0}	 & 	 326100.\phantom{0}	\\			
F$_3$ & 	4221000.\phantom{0}	 & 	1585000.\phantom{0}	 & 	 735000.\phantom{0}	\\			
\hline
\end{tabular}
}
\normalsize
\rm
\caption{ Lines from 2s$^2$ and 3s$^2$ ions. The first column gives the type of line, as defined in Fig. 1.6.}\label{t1}
\end{center}
\end{table}

%===============================================================================
\clearpage

   \vspace{20mm}
\noindent
{\bf Acknowledgements}
I thank Jordi Cepa for having invited me to share my experience on emission lines with the sparticipants to the XVIII Canary Island Winterschool ``The Emission line Universe'', and I thank the participants for their interest.
Much of the material presented here has benefited from discussions over the recent years with my colleagues, collaborators  and friends: Fabio Bresolin, Miguel Cervi\~no, Roberto Cid-Fernandes, Barbara Ercolano, C\'esar Esteban, Gary Ferland, Jorge Garc\'ia-Rojas, Rosa Gonz\'alez Delgado, S\l{}awomir G\'orny, William Henney, Artemio Herrero, Yuri Izotov, Luc Jamet, Valentina Luridiana, Abilio Mateus, Christophe Morisset, Antonio Peimbert, Manuel Peimbert, Miriam Pe\~na, Enrique P\'erez, Michael Richer, M\'onica Rodr\'iguez, Daniel Schaerer, Sergio Sim\'on-D\'iaz, Laerte Sodr\'e, Ryszard Szczerba, Guillermo Tenorio-Tagle, Silvia Torres-Peimbert, Yiannis Tsamis, Jos\'e V\'ilchez. Special thanks to  Jorge Garc\'ia-Rojas, \'Angel L\'opez-S\'anchez, Sergio Sim\'on-D\'iaz, and Natalia Vale Asari for their detailed proof-reading and comments on this manuscript. I apologize to the many researchers in the field who did not find their names in the list of references. It is unfortunately impossible in such a short text to do justice to all the work that has contributed to our understanding of the subject.
While preparing this review, I have made extensive use of the Nasa ADS data base. I have also used the Atomic Line List
maintained  by Peter van Hoof, from which the tables in the appendix were extracted, with the great help of Sergio Sim\'on-D\'iaz. In the body of this contribution, I have indicated several web sites that I think might be of interest to the reader.

\end{document}